\documentclass[sn-standardnature]{sn-jnl}

\usepackage{adjustbox}

\usepackage{cleveref}

\jyear{2023}%

\theoremstyle{thmstyleone}%
\theoremstyle{thmstyletwo}%

\theoremstyle{thmstylethree}%

\def\modelName{VOLTA}
\def\ovarianDatasetName{Oracle} 

\begin{document}

\title[\modelName]{\modelName: an Environment-Aware Contrastive Cell Representation Learning for Histopathology}


\author[1]{\fnm{Ramin} \sur{Nakhli}}
\author[2]{\fnm{Allen} \sur{Zhang}}
\author[1]{\fnm{Hossein} \sur{Farahani}}
\author[3]{\fnm{Amirali} \sur{Darbandsari}}
\author[1]{\fnm{Elahe} \sur{Shenasa}}
\author[4]{\fnm{Sidney} \sur{Thiessen}}
\author[4]{\fnm{Katy} \sur{Milne}}
\author[5]{\fnm{Jessica} \sur{McAlpine}}
\author[5]{\fnm{Brad} \sur{Nelson}}
\author[2]{\fnm{C Blake} \sur{Gilks}}
\author*[1,2]{\fnm{Ali} \sur{Bashashati}}\email{ali.bashashati@ubc.ca}


\affil[1]{\orgdiv{School of Biomedical Engineering}, \orgname{University of British Columbia}, \orgaddress{\city{Vancouver}, \country{Canada}}}

\affil[2]{\orgdiv{Department of Pathology and Laboratory Medicine}, \orgname{University of British Columbia}, \orgaddress{\city{Vancouver}, \country{Canada}}}

\affil[3]{\orgdiv{Department of Electrical Engineering}, \orgname{University of British Columbia}, \orgaddress{\city{Vancouver}, \country{Canada}}}

\affil[4]{\orgdiv{Deeley Research Centre}, \orgname{BC Cancer Agency}, \orgaddress{\city{Victoria}, \country{Canada}}}

\affil[5]{\orgdiv{Department of Gynecology and Obstetrics}, \orgname{University of British Columbia}, \orgaddress{\city{Vancouver}, \country{Canada}}}



\abstract{In clinical practice, many diagnosis tasks rely on the identification of cells in histopathology images. While supervised machine learning techniques require labels, providing manual cell annotations is time-consuming due to the large number of cells. In this paper, we propose a self-supervised framework (\modelName) for cell representation learning in histopathology images using a novel technique that accounts for the cell's mutual relationship with its environment for improved cell representations. We subjected our model to extensive experiments on the data collected from multiple institutions around the world comprising of over 700,000 cells, four cancer types, and cell types ranging from three to six categories for each dataset. The results show that our model outperforms the state-of-the-art models in cell representation learning. To showcase the potential power of our proposed framework, we applied \modelName\ to ovarian and endometrial cancers with very small sample sizes (10-20 samples) and demonstrated that our cell representations can be utilized to identify the known histotypes of ovarian cancer and provide novel insights that link histopathology and molecular subtypes of endometrial cancer. Unlike supervised deep learning models that require large sample sizes for training, we provide a framework that can empower new discoveries without any annotation data in situations where sample sizes are limited.}





\maketitle

\section{Introduction}\label{sec1}

Cells located within the micro-environment of a tumor have a prominent impact on its developmental process \cite{farc2021overview,yang2020tumor,liu2018transcriptome,bremnes2016role,schalper2015objective}. Variations in the micro-environment can influence the epigenetic profiles within the tumor and the heterogeneity in the associated gene expression profiles \cite{pogrebniak2018harnessing}. Various cell types reside in the tumor microenvironment and growing evidence suggest that this intratumoral heterogeneity vastly contributes to the therapeutic resistance of the tumor \cite{zhang2022tumor,pogrebniak2018harnessing}. Several studies have shown that the higher levels of intratumoral heterogeneity are strongly associated with poor outcomes in lung, ovarian, head and neck, and pancreatic cancers, as it implies that the tumor is more likely to  harbor a rare pre-existing resistant subclone \cite{pogrebniak2018harnessing,zhang2014intratumor,schwarz2015spatial,andor2016pan}.
Furthermore, the spatial distribution of immune cells within the tumor microenvironment has a significant impact on the prognosis and therapeutic responses \cite{steinhart2021spatial,fu2021spatial,bremnes2016role,brambilla2016prognostic,corredor2019spatial}. Therefore, the identification of individual cells within the tumor microenvironment is a vital step for tumor characterization in many complex tasks such as tissue classification, cancer diagnosis, subtyping and histological grading \cite{javed2020cellular,zhou2019cgc,martin2021predictive,lu2020capturing}. 

The visual assessment of the Hematoxylin \& Eosin (H\&E)-stained tissue slides under the microscope is the conventional and widely utilized approach to tumor characterization and cell identification. However, manual cell identification can be cumbersome due to both the time-consuming nature of assessment of large numbers of cells (tens of thousands in a single slide) and the intra- and inter-observer variability \cite{elmore2015diagnostic}. Machine learning and deep learning models coupled with the digitization of pathological material offer opportunities for computer-aided cell identification \cite{sirinukunwattana2016locality,graham2019hover,amgad2021nucls}. Despite the long history of machine learning research in cell classification using handcrafted features \cite{dalle2009nuclear,nguyen2011prostate,cruz2013deep}, significant improvements have been reported by solely employing the deep learning-based models \cite{graham2019hover}.

Even though these models can potentially reduce the manual workload of cell identification, they require a large number of cell-level annotations for training. However, this annotation collection process still relies on labor-intensive manual examination of the tissue by pathologists. Furthermore, to apply these models to a new tissue type, the data collection and labeling process has to be carried out again. 
To address this issue, a number of studies have utilized unsupervised approaches for cell representation learning and clustering. Hu et al. \cite{hu2018unsupervised} adopt InfoGAN \cite{chen2016infogan} to train an implicit classifier, and in another attempt, Vunuu et al. \cite{vununu2020strictly} use a deep convolutional auto-encoder (DCAE) to learn the embeddings of cells. However, these studies focus on only one tissue type and also ignore the surrounding environment of a cell while many studies have shown that cells are directly impacted by their environment \cite{keren2018structured,schurch2020coordinated,yuan2016spatial}. The former jeopardizes the generalization of the models from one tissue to another while the latter can potentially have an impact on the accuracy of the model.

Recently, self-supervised learning (SSL) techniques have emerged as an important step towards generalizable representation learning. SSL is a technique developed for image representation learning that guides the training of the model by using the augmentations of an image as the label for that image. The utility of this technique has been investigated on different tasks in the natural image domain where Caron et al. \cite{caron2021emerging} demonstrate the capability of this technique in object classification, and Sohn et al. \cite{sohn2020simple} show its efficacy in object detection. Despite the fact that the a few studies \cite{ciga2022self,zhang2021histopathology} examine the utility of self-supervised methods in the patch-level classification, the potential of self-supervised techniques for labeling individual cells (rather than just classifying image patches) are largely ignored. Furthermore, it is difficult to link the findings of these studies to the biological mechanisms in the tumor micro-environment as they cannot identify individual cells.

In this paper, we propose a self-supervised framework for cell representation learning in histopathology images by introducing a novel technique to incorporate the mutual relationship between the cell and its environment for improved cell representation. We benchmarked our model on data representing more than $700,000$ cells in four cancer histotypes with three to six cell types in each dataset. Results confirm the superiority of our model in memory-efficient cell type representation compared to the state-of-the-art. We further utilized the proposed model in the context of ovarian and endometrial cancers and demonstrated that our cell representations, without any human annotations, can be utilized to identify the known histotypes of ovarian cancer  as well as identifying novel insights that link histopathology and molecular subtypes of endometrial cancer.

\section{Methods}\label{sec:method}

\begin{figure}[t]
    \centering
    \includegraphics[width=\textwidth]{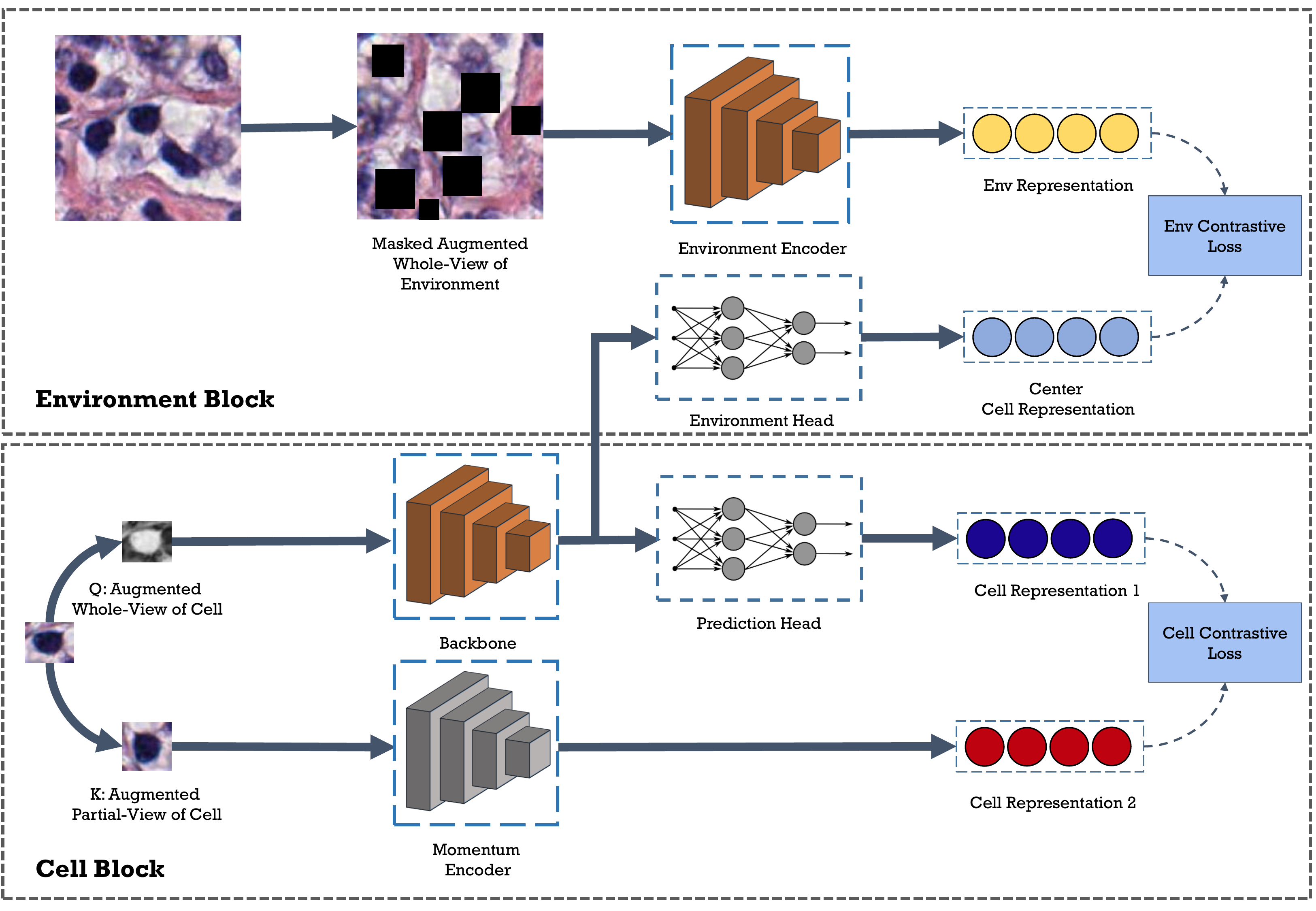}
    \caption{Overview of our proposed framework. The cell block trains the backbone model by applying two augmentations on the same cell image, encoding the images, and bringing their representations close to each other. While the backbone is trained through back-propagation, the momentum encoder averages the weights from the backbone. On the other hand, the environment block combines the cell representation created by the cell block with the surrounding environment (a larger region around the cell). We mask all of the cells in the environment patch to prevent the model from favoring the cell representation toward that of these cells.} 
    \label{overview_fg}
\end{figure}

\Cref{overview_fg} depicts an overview of the proposed self-supervised method for cell classification. In this figure, $X$ is the input cell image, $Q$ together with $K$ are the augmented cell images, $E$ is the environment image, $E_{Aug}$ is the augmented environment image, and $L^{cell}$ alongside $L^{Env}$ show the cell and environment loss values. \par

This framework consists of two main blocks: 1) Cell Block; 2) Environment Block. The Cell Block learns the cell embeddings by contrasting individual cell-level images while the Environment Block incorporates environment-level information into the cell representations. \par 

\subsection{Cell Block}

The architectural design of the Cell Block is similar to our previously proposed model~\cite{nakhli2022ccrl}, which has shown promising performance in cell representation learning tasks. In this block, cell embeddings are learned by pulling the embeddings of two augmentations of the same image together, while the embeddings of other images are pushed away. Let $X=\{x_i \mid 1\leq i\leq N\}$ be the input batch of cell images and $N$ to be the number of images in the batch. Each $x_i$ is a small crop of the H\&E image around a cell in a way that it only includes that specific cell. Two different sets of augmentations are applied to $X$ to generate $Q=\{q_i \mid 1\leq i\leq N\}$ and $K=\{k_i \mid 1\leq i\leq N\}$. We call these sets query and key, respectively. $q_i$ and $k_j$ are the augmentations of the same image if and only if $i=j$. The query batch is encoded using a backbone model, a neural network of choice, while the keys are encoded using a momentum encoder, which has the same architecture as the backbone. This momentum encoder is updated using \Cref{moemtum_eq} in which $\theta_{k}^{t}$ is the parameter of momentum encoder at time $t$ ,$m$ is the momentum factor, and $\theta_{q}^{t}$ is the parameter of the backbone at time $t$

\begin{equation} \label{moemtum_eq}
    \theta_{k}^{t} = m\theta_{k}^{t-1} + (1-m)\theta_{q}^{t}.
\end{equation}

Consequently, the obtained query and key representations are passed through separate Multi-Layer Perceptron (MLP) layers called projector heads. Although the query projector head is trainable, the key projector head is updated with momentum using the weight of the query projector head. We restrict these layers to be 2-layer MLPs with an input size of 512, a hidden size of 128, and an output size of 64. In addition to the projector head, we use an extra MLP on the query side of the framework, called the prediction head. This network is a 2-layer MLP with input, hidden, and output sizes of 64, 32, and 64, respectively. Similar to the last fully-connected layers of a conventional classification network, the projection and prediction heads provide more representation power to the model. \par

The networks of the cell block are trained using the InfoNCE \cite{oord2018representation} loss which is shown in \Cref{infonce_eq}

\begin{equation} \label{infonce_eq}
    L_{q_{i}}^{cell} = -\log \frac{\exp\frac{\Vert f_q(q_i) \Vert^2 . \Vert f_k(k_i) \Vert^2}{\tau}}{\sum_{j=0}^{N+Q} \exp\frac{\Vert f_{q}(q_i) \Vert^2 . \Vert f_k(k_j) \Vert^2}{\tau}}
.\end{equation}

In this equation, $\tau$ is the temperature that controls the sharpness of the distribution, $\Vert \Vert$ is the cardinality, $Q$ is the number of items stored in the queue from the key branch, $f_q$ is the equal function for the combination of the backbone, query projection head, and query prediction head, and $f_k$ shows the equal function for the momentum encoder and the key projection head. \par

The augmentation pipelines include cropping, color jitter (brightness of 0.4, contrast of 0.4, saturation of 0.4, and hue of 0.1), gray-scale conversion, Gaussian blur (with a random sigma between 0.1 and 2.0), horizontal and vertical flip, and rotation (randomly selected between 0 to 180 degrees). To ensure the model always views the whole-cell image of the cell on one side, we remove the cropping operation from one of the pipelines. Therefore, the pipeline with cropping generates local regions of the cell image while the images generated by the other augmentation pipeline are global, containing the whole-cell view. \par

Cell embeddings are generated from the trained momentum encoder at the inference time and are clustered by applying the K-means algorithm. One can use either the encoder or momentum encoder for embedding generation; however, the momentum encoder provides more robust representations since it aggregates the learned weights of the encoder network from all of the training steps (an ensembling version of the encoder throughout training)~\cite{caron2021emerging}. \par

\subsection{Environment Block}

Many studies have shown that the Tumor Micro Environment (TME) plays an important role in the tumor progression behavior~\cite{blise2022single,yuan2016spatial}. Motivated by these findings, we ask: should the representation of a cell reflect its environment as well? Inspired by this question, we hypothesize that a deeper knowledge of the environment leads to a better general understanding of the cell. In a mathematical formulation, this hypothesis is equivalent to the assumption that there exists mutual information between cells and their environment. Therefore, to validate this hypothesis, we propose to increase the mutual information between the corresponding cell and environment representations during the training process. Previous studies~\cite{wu2020mutual} have shown that the InfoNCE loss maximizes the lower bound of mutual information between different views of the image. Thus, we will use this loss function to achieve the aforementioned target by performing cross-modal contrastive learning as an auxiliary task. \par

Let $E=\{e_i \mid 1\leq i\leq N\}$ be the corresponding environment patches of the cells represented by $X$. Here, we refer to the environment as a large region around a cell in a way that includes the surrounding tissue and cells. Therefore, for $\forall i\in 1,2,..., N$, $x_i$ and $e_i$ are centered on the same cell (however, for the cases where the cells are located on the edge of the patch, we limit the patch border to the border of the image). After applying an augmentation pipeline, the environment patches are passed through an encoder network, called an environment encoder. Simultaneously, we apply a new projection head, the environment projection head, to the cell representations obtained from the query backbone in the Cell Block. Finally, one can train the Environment Block using these two sets of representations (environment and cell) and \Cref{env_loss_eq}

\begin{equation} \label{env_loss_eq}
    L_{q_{i}}^{env} = -\log \frac{\exp\frac{\Vert g_{cell}(q_i) \Vert^2 . \Vert g_{env}(e_i) \Vert^2}{\tau}}{\sum_{j=0}^{N} \exp\frac{\Vert g_{cell}(q_i) \Vert^2 . \Vert g_{env}(e_j) \Vert^2}{\tau}}.
\end{equation}

Therefore, the final loss of the whole framework can be written as \Cref{total_loss_eq}, in which $\lambda$ is a hyperparameter. Increasing the value of $\lambda$ prioritizes the mutual information of the cell with its environment over the consistency of the representation for different augmentations of the same cell

\begin{equation} \label{total_loss_eq}
    L_{q_i} = L^{cell}_{q_i} + \lambda L^{env}_{q_i}.
\end{equation}

The augmentation pipeline of the Environment Block uses the same operations as that of the Cell Block except for cropping. \par

To prevent the model from focusing on the overlapping regions between the corresponding cell and environment images (called shortcut~\cite{minderer2020automatic}, meaning that the model uses undesired features to solve the problem), we mask the target cell in the environment patch. Furthermore, the rest of the cells in the environment patch are also masked to ensure that the model does not bias the representation of a cell toward the neighboring cell types. We will investigate the effectiveness of the masking operation in the ablation study. \par

\subsection{Data Preparation}

The aforementioned datasets included patch-level images, while we required cell-level ones for the training of the model. To generate such data, we used the instance segmentation provided in each of the external datasets to find cells and crop a small box around them. However, for the \ovarianDatasetName and SarcCell datasets, the instance segmentation masks were generated by applying HoVer-Net~\cite{graham2019hover} segmentation pre-trained on the PanNuke dataset. \par

An adaptive window size was used to extract cell images from the H\&E slides. The window size was set to twice the size of the cell for the CoNSeP dataset while it was equal to the size of the cell for the rest of the datasets. Finally, cell images were resized to $32\times 32$ pixels and were normalized to zero mean and unit standard deviation before being fed into our proposed framework. \par

Ground-truth label generation of the \ovarianDatasetName and SarcCell dataset cells was performed by finding the most expressed biomarker (by intensity and quantity) in the same position of the corresponding IHC image. To accommodate for the potential noise associated with image registration, two post-processing steps were performed: 1) the size of the window in the IHC image was set to 5 times of the window size in the H\&E core (however, this scale was set to 1 for the SarcCell dataset due to more accurate co-registration performance); 2) the most expressed biomarker was considered as the label only if it contained at least 70\% of the biomarker distribution in the IHC window.

\subsection{Implementation Details}
The code was implemented in Pytorch (v1.9.0), and the model was run on one and two V100 GPUs for the w/ and w/o environment settings, respectively. The batch size was set to 1024 (unless specified otherwise), the queue size to 65536, and pre-activated ResNet18~\cite{he2016identity} was used for the backbone and momentum encoder in the Cell Block. The environment encoder architecture was set to LambdaNet model~\cite{bello2021lambdanetworks} as it extracts more informative patch representations using self-attention while keeping the computation and memory usage tractable. The stack was trained using the Adam optimizer for 500 epochs (unless specified otherwise) with a starting learning rate of 0.001, a cosine learning rate scheduler, and a weight decay of 0.0001. We also adopted a 10-epoch warm-up step. The momentum factor in the momentum encoders was 0.999, and the temperature was set to 0.07. \par

In \Cref{results_tb} experiments, the training epoch count and batch size of our models were set to 200 and 512 for the PanNuke Breast, Lizard, \ovarianDatasetName, and SarcCell datasets. Additionally, for the training of our model on the \ovarianDatasetName datasets, we used 15,000 randomly selected cells from the training set, to reduce the training time.

In the self-supervised to supervised transfer learning step (cell classification), we adopted SGD (Stochastic Gradient Descent) with a starting learning rate of 0.001 using a cosine learning rate scheduler for 300 epochs with a batch size of 1024. Also, the weight decay was set to 0.00001. In the case that we allowed the encoder to be fine-tuned, we set the encoder’s learning rate to 0.0001. \par

It is worth mentioning that for the cell classification of NuCLS, we followed the same super-class grouping of the original paper~\cite{amgad2021nucls}. In this regard, we only used 3 super-classes out of 5 for cell type classification, including tumor, stromal, and sTILs. \par

\subsection{Baselines}
The performance was also compared against five baselines. The pre-trained ImageNet model used weights that were pre-trained on the ImageNet dataset to generate the cell embeddings. The Morphological Features approach ~\cite{van2008visualizing} adopted morphological features to produce a 30-dimensional feature vector, consisting of geometrical and shape attributes. Prior to clustering, the feature vectors were normalized to zero mean and unit standard deviation, and their size was reduced to 2 using t-SNE. The third baseline was Manual Feature~\cite{hu2018unsupervised} which used a combination of Scale-Invariant Feature Transform (SIFT) and Local Binary Patterns (LBP) features to provide representations for the cells. Similar to the previous baseline, we exercised standardization on the computed feature vectors. Additionally, our baseline set included two state-of-the-art unsupervised deep learning models. More specifically, the Auto-Encoder baseline adopted a deep convolution auto-encoder alongside a clustering layer to learn cell embeddings by performing an image reconstruction task~\cite{vununu2020strictly}. And finally, the last baseline was GAN~\cite{hu2018unsupervised} which adopted the idea of InfoGAN~\cite{chen2016infogan} and developed a Generative Adversarial Network (GAN) for cell clustering by increasing the mutual information between the cell representation and a categorical noise vector. 

\section{Results}\label{sec2}

\subsection{Cell representation learning framework and benchmarking}
\label{sec:results_intro}
\Cref{overview_fg} depicts an overview of our proposed Environment-Aware Contrastive Cell Representation Learning framework (\modelName). This framework consists of two major blocks. The \textit{cell block} takes an image of a cell and applies two sets of augmentation operations to create visually distinct perspectives of a cell. These two augmented images are then transformed into their respective representation vectors using a stack of deep neural networks and given that these representations correspond to the same cell, the models are trained to minimize the distance between the two representations. The \textit{environment block} of our proposed framework is utilized to increase the mutual information between the cell and a larger patch that captures the environment around it. By using the InfoNCE loss function \cite{oord2018representation}, it accomplishes this by performing a contrastive cross-modal learning between the cell representation and that of its environment. To prevent the model from biasing towards other cells appearing in the environment, we mask out these cells in the environment patch before feeding it to the model. Finally, the cell representations for downstream tasks such as cell clustering and classification can be obtained by using the backbone model trained in this setting. 
 
We benchmarked these representations across multiple tasks and datasets. More specifically, seven public and private datasets representing $700,000$ cells and four cancer types (colon, breast, and ovarian cancers and sarcomas) were utilized to evaluate the performance of the proposed cell representation model (Supplementary \Cref{dataset_tb}). Even though our model requires no labels for training, in all settings, we split the data into train and test sets and use the former for the training of the model. 

We also conducted ablation studies on different components of our model to measure their effects on the performance (see Supplementary \Cref{sec:ablation_study}). Our experiments suggest that the cell masking operation, whole- and local-view augmentations, memory storage, and momentum encoder provide noticeable performance improvements to our model.

\subsection{Identification of distinct cell clusters by self-supervised cell representation learning}

\begin{table}
\begin{center}
\caption{Unsupervised clustering of cell representations across different methods and datasets. The baseline models include both morphology-based and state-of-the-art deep learning methods for cell representation. Some of the baseline results are listed as "-" meaning calculation of the feature vectors was not possible due to the limitation of the model on the small-sized cells.}\label{results_tb}
\adjustbox{max width=\textwidth}{
\begin{tabular}{@{\extracolsep{\fill}}lccccccccc@{\extracolsep{\fill}}}
\toprule
Model                                   & Metric & CoNSeP          & NuCLS           & PanNuke Breast   & PanNuke Colon     & Lizard            & \ovarianDatasetName               & SarcCell        \\
\midrule
\multirow{3}{*}{Pre-trained ImageNet}   & AMI    & 7.3\%           & 9.3\%           & 5.42\%           & 11.21\%           & 6.25\%            & 0.26\%           & 4.7\%          \\ 
                                        & ARI    & 7\%             & 7.8\%           & 3.94\%           & 8.21\%            & 4.36\%            & 0.42\%           & 4.2\%          \\ 
                                        & Purity & 42.7\%          & 56.7\%          & 41.15\%          & 43.93\%           & 50.4\%            & 48.87\%          & 55.1\%         \\ 
\midrule
\multirow{3}{*}{Morphological}          & AMI    & 12.7\%          & 21.1\%          & 8.94\%           & 7.88\%            & 13.21\%           & -                & -              \\ 
                                        & ARI    & 1.3\%           & 18.8\%          & 7.28\%           & 6.19\%            & 9.22\%            & -                & -              \\ 
                                        & Purity & 48.8\%          & 66.1\%          & 47.06\%          & 42.73\%           & 57.5\%            & -                & -              \\
\midrule
\multirow{3}{*}{Manual Features}        & AMI    & 9.5\%           & 11.25\%         & -                & 7.86\%            & 10.2\%            & 2.74\%           & 1.6\%          \\ 
                                        & ARI    & 6.4\%           & 7.8\%           & -                & 6.53\%            & 3.8\%             & 2.24\%           & 1.5\%          \\ 
                                        & Purity & 45.5\%          & 56.2\%          & -                & 40.37\%           & 52.9\%            & 53.84\%          & 51.1\%         \\
\midrule
\multirow{3}{*}{DCAE}                  & AMI    & 10.1\%          & 8.3\%           & 6.41\%           & 11.43\%           & 4.36\%            & 3.93\%           & 4.0\%          \\ 
                                        & ARI    & 7.3\%           & 7.2\%           & 5.11\%           & 10.01\%           & 2.34\%            & 3.84\%           & 6.1\%          \\ 
                                        & Purity & 50.5\%          & 56.8\%          & 43.49\%          & 45.18\%           & 49.38\%           & 58.69\%          & 57.1\%         \\
\midrule
\multirow{3}{*}{GAN}                    & AMI    & 14.8\%          & 14\%            & 6.7\%            & 13.7\%            & 7.5\%             & 4.1\%            & 4.8\%          \\ 
                                        & ARI    & 15.7\%          & 12.6\%          & 4.6\%            & 11.4\%            & 3\%               & \textbf{5.8\%}   & 6.2\%         \\ 
                                        & Purity & 58.4\%          & 62\%            & 42.4\%           & 49.6\%            & 48.9\%            & 57.5\%           & 58.4\%          \\
\midrule
\multirow{3}{*}{\modelName}     & AMI    & \textbf{25.5\%} & \textbf{26.2\%} & \textbf{13.8\%}  & \textbf{22.5\%}   & \textbf{17.3\%}    & \textbf{8.05\%}  & \textbf{7.9\%} \\ 
                                        & ARI    & \textbf{19.3\%}          & \textbf{27.3\%} & \textbf{8.94\%}  & \textbf{21.8\%}   & \textbf{11.4\%}    & 4.95\%           & \textbf{9.7\%} \\ 
                                        & Purity & \textbf{63.5\%} & \textbf{70.3\%} & \textbf{47.7\%}  & \textbf{56.9\%}   & \textbf{57.9\%}    & \textbf{59.45\%} & \textbf{60.3\%}\\
\bottomrule
\end{tabular}
}
\end{center}
\end{table}

\modelName\ provides cell representations from histopathology images, and such representations should be capable of differentiating between biologically distinct cell types. To test this hypothesis, we adopted our method to identify cell clusters in each dataset. In particular, after learning the cell representations in a self-supervised manner using \modelName, we performed unsupervised clustering on the cell representations and examined the enrichment of the identified clusters with specific cell types. To show the utility of our approach, we compared the performance of \modelName\ with the state-of-the-art morphology-based and deep learning-based models for cell representation. As shown in \Cref{results_tb}, our model outperformed all counterparts by a large margin across multiple clustering metrics (adjusted mutual index (AMI), adjusted rand index (ARI), and Purity - see Supplementary \Cref{sec:evaluation_metrics}) in all datasets, reaching twice the performance of the best-performing baselines in some of the datasets (except for ARI on the Oracle dataset where GAN performs better). More importantly, while the performance of the baseline models varies from one cancer to another, our model shows consistent results regardless of the cancer type. For instance, while the morphology-based representation method has the best performance compared to the other baselines over the NuCLS and PanNuke Breast cancer datasets, it has an inferior performance on PanNuke Colon and CoNSeP.

\begin{figure}[t]
    \centering
    \includegraphics[width=\textwidth]{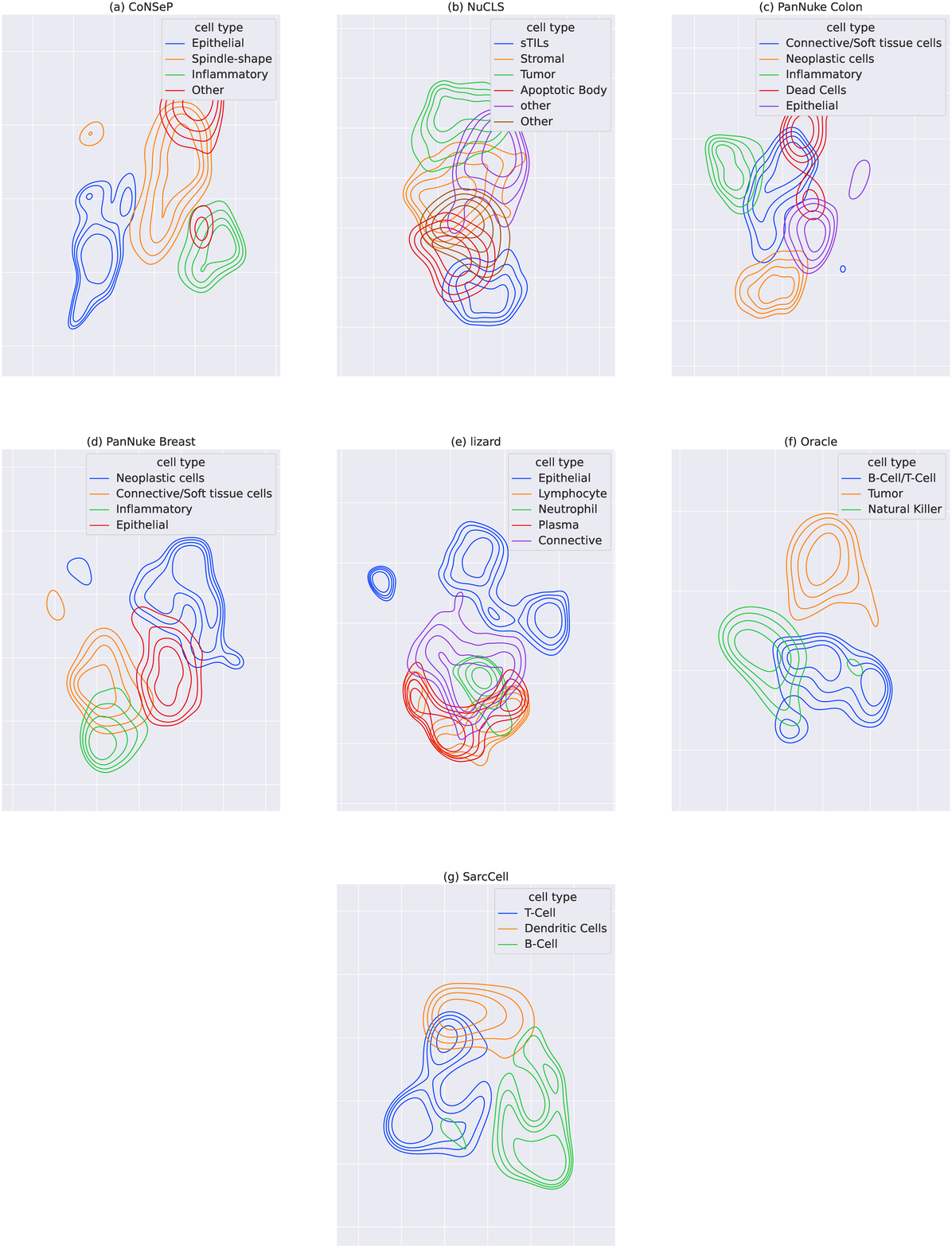}
    \caption{Embedding space representation of each dataset using UMAP. (a) CoNSeP, (b) NuCLS, (c) PanNuke Colon, (d) PanNuke Breast, (e) Lizard, (f) \ovarianDatasetName, (g) SarcCell. Contours with the same color demonstrate the distribution of the learned representations by our model for that specific cell types. Despite not using labeled data in the training process, our model learns to map cells with the same type close to each other. The co-centered contours with the same color show the distribution of the representation for cells with a specific type. } 
    \label{embedding_fg}
\end{figure}

\Cref{embedding_fg} shows the Uniform Manifold Approximation and Projection (UMAP) representations of various cell types that were derived by \modelName. The learned representations provide distinct and separable cell populations confirming the comparison metrics that were presented in \Cref{results_tb}. Also, one can observe that our model can differentiate between immune cells (T-cell, B-cell, and Natural Killer cells) and tumor cells in the \ovarianDatasetName\ dataset. Similarly, in the NuCLS dataset, our model is able to differentiate between the stromal tumor-infiltrating lymphocytes (sTILs) and the cancer cells. The same observations can be seen in the PanNuke Colon and CoNSeP datasets where various cell types such as epithelial and inflammatory cells are mapped onto different locations of the embedding space.

\subsection{Supervised cell classification accuracy and efficiency improvement}

\begin{figure}[t]
    \centering
    \includegraphics[width=\textwidth]{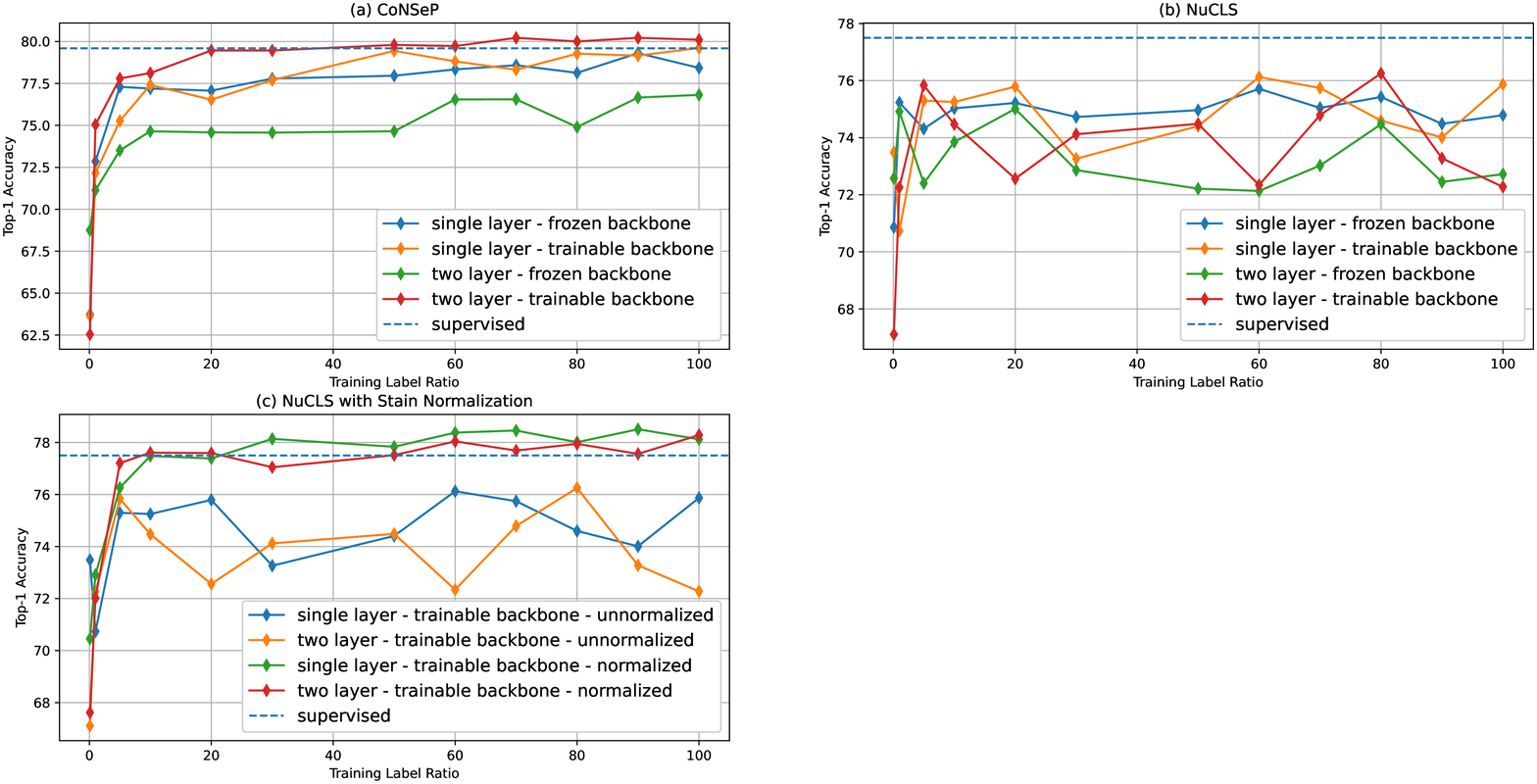}
    \caption{After pre-training using our self-supervised framework, a fully-connected layer (single- or double-layer) was added to the end of the backbone (the model generating the cell representations), and they were fine-tuned using the labeled data. We compared fine-tuning with both frozen and unfrozen backbone (a - CoNSeP and b - NuCLS). To account for the color differences in the train and test cohorts of the NuCLS dataset, we also performed the Vahedain color normalization before the fine-tuning process, which showed a significant boost compared to the unnormalized approach (c). The results demonstrate that our fine-tuned model can achieve the same performance as the supervised baselines (HoVer-Net and NuCLS) using only 20\% of the labeled data while outperforming these baselines with the full set of the labeled data (a and c).} 
    \label{finetune_fig}
\end{figure}

We then aimed to assess the effectiveness of the proposed model in few-shot cell classification in a supervised machine learning setting where labeled samples were available. Specifically, we trained the model using our self-supervised framework and utilized the learned cell representations as inputs for training a simple Multi-Layer Perceptron (MLP) for cell classification. The performance of the trained model on CoNSeP and NuCLS datasets across various settings is shown in Figure \ref{finetune_fig}. 

We also demonstrated the effectiveness of our self-supervised cell representation learning framework by using a subset of the labeled cell identities to train the MLP-based cell classifier. Our results showed that the proposed model achieved a reasonable performance with a small subset of the labeled training data (Supplementary \Cref{finetuning_tb}). For instance, with only 0.1\% of the training labels, our models achieved 62.7\% and 72.6\% Top-1 accuracy on the CoNSeP and NuCLS datasets, respectively, while a model that utilized the entire labeled dataset achieved 80.2\% and 76.3\%. Furthermore, as the number of training labels increased, the classification accuracy consistently improved to an extent that our model outperformed state-of-the-art Hover-Net model \cite{graham2019hover} results on the CoNSeP dataset, even with 70\% of the training data. It is of note to mention that the number of the parameters of our proposed model is reduced by 60\% compared to the HoVer-Net model (Supplementary \Cref{model_size_tb}). Our model reached an accuracy that was close to the Masked-RCNN model which led to state-of-the-art results in the NuCLS dataset \cite{amgad2021nucls}. Given that the training and validation sets of this dataset are collected from different sites, we hypothesize that variations in staining and color profiles could lead to the over-fitting of the models to training data (see \Cref{sec:color_normalization}).

\subsection{Self-supervised cell representation learning is robust to undesired color variations}
\label{sec:color_normalization}
Previous studies have shown that normalization and domain adaptation methods can enhance the performance of supervised models when the train and test datasets are collected from different sites. Therefore, we studied the effect of such methods on our proposed model when it was utilized for cell representation learning and supervised cell classification settings. To serve this purpose, we used the Vahadane normalization method \cite{vahadane2016structure} within the context of the NuCLS dataset where the slides were stained and scanned in different institutions. \par 

Supplementary \Cref{normlization_tb} illustrates the effect of the normalization in the self-supervised setting on the NuCLS dataset. Although patch and slide classification tasks can benefit from cross-institution stain normalization, we noticed that our self-supervised cell representation approach does not benefit much from color normalization strategies. This finding can be attributed to the strong augmentations that were utilized in our self-supervised model training. Moreover, we investigated the effect of color normalization in the supervised fine-tuning setting. Interestingly, although self-supervised clustering results were robust to stain normalization, the supervised fine-tuned model benefited from it to an extent that it outperformed the NuCLS model on this dataset (Supplementary \Cref{nucls_finetuning_tb}). It is of note to mention that the normalization method was only applied to the test set while the self-supervised model was still trained on the original data (i.e., without any normalization).

\subsection{\modelName\ as a building block for unsupervised cancer subtype identification}

\begin{figure}[t]
    \centering
    \includegraphics[width=\textwidth]{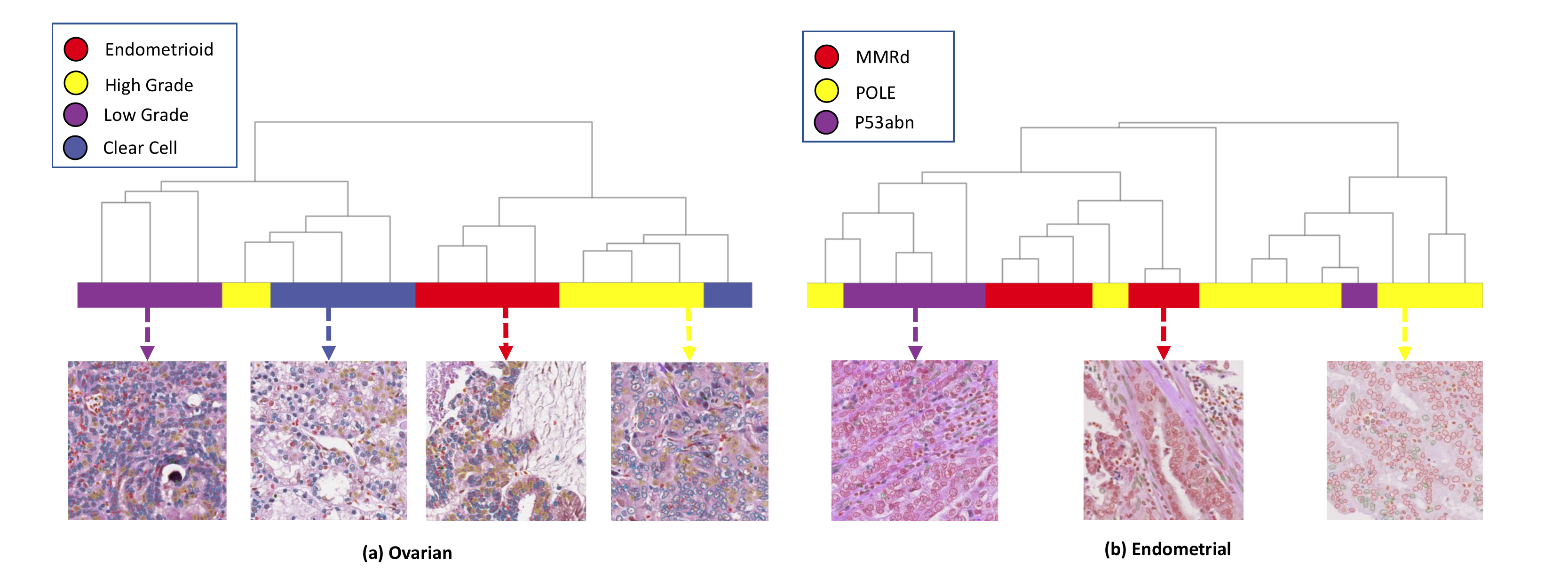}
    \caption{(a) Ovarian cancer and (b) endometrial cancer datasets are hierarchically clustered based on cell cluster proportions. To achieve this, we first train our model for delivering cell representations in a self-supervised manner. Then, it will be applied to patches of a slide, the cell cluster distributions will be counted, and the slides will be clustered into distinct cohorts based on the variation of the cell cluster distributions across patches of each slide. In the case of endometrial cancer (b), the supercluster on the right (yellow) demonstrates a cohort of patients that mostly have the POLE subtype (only one sample from p53abn is in this group), the supercluster in the middle (red) depicts mainly the MMRd patients (with only one POLE case misclassified), and the superclass on the left (purple) shows the p53abn cases with only one POLE case misplaced.} 
    \label{cancer_subtype_clustering}
\end{figure}

We sought to investigate the utility of our proposed self-supervised cell representation model as a building block for annotation-free cancer subtyping. Therefore, we put together a small H\&E tissue microarray (TMA) cohort of 14 ovarian cancer cases comprising of clear cell, endometrioid, high-grade serous, and low-grade serous ovarian carcinomas. By exercising the same procedure as described in \Cref{sec:results_intro}, we utilized the cells extracted from these images to train our self-supervised model. Subsequently, TMA core images were divided into $400\times400$ pixel patches, our pre-trained \modelName\ model was applied to each patch to derive cell representations, cells were divided into multiple clusters, and the number of cells within the predicted clusters was counted for each patch. To prevent biasing towards patches that appear frequently (e.g., stromal regions), we used a clustering method to split all of the patches into multiple distinct clusters based on the aforementioned cell distributions. Afterwards, we selected 100 patches from each cluster and used their cell distributions to predict the clustering label of the original TMA core (see Supplementary \Cref{cancer_subtyping_workflow_fg}). The results demonstrate that our model is capable of separating the epithelial ovarian cancer histotypes without a need for annotation or prior knowledge of the histotypes (\Cref{cancer_subtype_clustering}a). In particular, four major clusters enriched with each of the four specific histotypes were identified with only two cases that were grouped with other subtypes. These results suggest an 86\% accuracy (12 out of 14 that were correctly grouped) in ovarian cancer subtyping; a finding that is in line with results reported in the literature \cite{farahani2022deep}. To make this process more interpretable, we visualized the cell clusters on multiple patches, asked an expert to label each cluster, and combined the clusters with the same label (referred to as sub-clusters). We observed that each of the cell clusters is typically enriched with a specific type of cell, demonstrating the capability of the model in capturing morphological differences between cell types (\Cref{ovarian_clear_overlay_fg,ovarian_mucinous_overlay_fg,ovarian_low_grade_overlay_fg,ovarian_high_grade_overlay_fg,ovarian_endometriod_overlay_fg}). Supplementary \Cref{ovarian_cluster_tb} represents the cell distributions across the epithelial ovarian histotypes after combining sub-clusters, while Supplementary \Cref{ovarian_boxplot_fg} depicts the boxplot of the cell distributions before combination. Notably, we noticed that 5 identified sub-clusters corresponded to differences in tumor cell morphology between ovarian cancer histotypes. High-grade serous and clear cell tumors were relatively enriched for tumor cell clusters containing larger cells (tumor clusters 2, 4, and 5) compared to low-grade serous and endometrioid tumors (see \Cref{fig:ovarian_cell_area,fig:ovarian_tumor_245}), consistent with the well-known high-grade nuclear histology of high-grade serous and clear cell carcinomas. 

We next utilized \modelName\ to demonstrate its application for exploratory cancer subtype discovery. To do so, we scanned 19 whole-section slide images (WSI) corresponding to three molecular subtypes of endometrial cancer (EC): DNA polymerase epsilon (POLE)-mutant cases, cases with mismatch repair deficiency (MMRd), and cases with p53 abnormality as assessed by immunohistochemistry. We next asked whether the proposed model could identify features in the H\&E slides that would aid us in identifying molecular subtypes of EC. Following a similar approach that we took for the ovarian cancer cohort, we subjected EC WSI representations to clustering and identified three clusters of patients (\Cref{cancer_subtype_clustering}b). Interestingly, each of the three clusters was enriched with a specific molecular subtype of endometrial carcinoma. 
 Similar to the procedure taken for the ovarian dataset, we also visualized the cell clusters within a patch for each of the EC subtypes (Supplementary \Cref{endometrial_pole_overlay_fg,endometrial_mmrd_overlay_fg,endometrial_p53abn_overlay_fg}). Furthermore, the distribution of cell clusters across the predicted histotypes and molecular subtypes before and after sub-cluster combination is shown in Supplementary \Cref{endometrial_cluster_tb} and \Cref{endometrial_boxplot_fg}, respectively. In line with recent findings, MMR-deficient tumors had the highest proportion of lymphocytes in the endometrial cancer dataset \cite{pasanen2020clinicopathological,ramchander2020distinct,dong2021pole}.

\section{Discussion \& Conclusion}\label{sec}
In this paper, we proposed a novel self-supervised framework (\modelName) for learning cell representations from annotation-free H\&E images. Our investigations from multiple perspectives confirm the superiority of \modelName\ over the state-of-the-art models. Specifically, we demonstrated that \modelName\ significantly outperformed the state-of-the-art unsupervised morphology- and deep-learning-based cell clustering methods on seven datasets, four cancer types, and three to six cell type categories within each dataset. Such unsupervised learning of the cell representations introduces unique opportunities for discovery, prediction, and development purposes. For instance, as part of our experiments, we illustrated that \modelName\ can be successfully used as a building block for cancer histotype clustering by applying it to 14 cases of ovarian and 19 cases of endometrial cancer, separately. This finding is interesting from two aspects: 1) even though our model does not receive any patient labels at training time, it is able to identify clusters of patients that are similar to pathologist diagnosis or molecular subtypes; 2) \modelName\ is extremely data efficient to an extent that it worked on two datasets with 10-20 patients samples while having a large dataset is usually a prerequisite for other deep learning models. We also demonstrated that these improvements are not only exclusive to the unsupervised aspects of the model but also can be extended to a supervised setting. By using our pre-trained \modelName\ as an initialization weight for a supervised model, we could achieve a performance equal to that of the state-of-the-art supervised model with as low as 10\% of the labeled data, while it surpassed this performance with the full data. Additionally, our self-supervised model is robust to undesired staining biases, which facilitates the utilization of the model on datasets collected across different centers.

Our investigation has demonstrated the efficiency of \modelName\ as a tool for cell discovery within multiple pathology pipelines. Leveraging a self-supervised engine, the model can be seamlessly integrated with a wealth of histopathology archives accessible from various clinical centers to enable the generation of extensive cell-level representation databases. Furthermore, the model has the potential to alleviate the laborious cell type labeling process by annotating cell clusters instead of individual cells and be used in an interactive pathology pipeline. In addition to its utilization in cell type discovery, we have also demonstrated that the model can serve as a foundational element for both histotype and molecular subtype identification. This illustrates the wide-ranging potential of our model for discovery at multiple levels, from morphological features to molecular basis. These findings point to interesting directions for linking histopathology data to more advanced and in-depth areas such as genomic and molecular information.

The spatial distribution of cells within a tumor has been widely acknowledged to have a profound impact on the progression and prognosis of the disease. As demonstrated by Pogrebniak et al. \cite{pogrebniak2018harnessing}, the bivariate analysis of immune and tumor cells can yield a wealth of information about the underlying biology of the disease. By utilizing metrics such as the Morisita–Horn index \cite{scalon2011spatial}, Ripley’s K function \cite{ripley1976second}, and Intra-Tumor Lymphocyte Ratio (ITLR) \cite{yuan2015modelling}, researchers have gained meaningful insights into the relationship between the spatial distribution of cells and clinical outcomes, identify immune-cancer hotspots, and predict chemotherapy response \cite{yuan2016spatial,denkert2010tumor,blise2022single}. Considering the crucial role of cell identification in these applications, our research has the potential to be instrumental in enabling the aforementioned studies to be conducted at more extensive scales. This, in turn, can lead to a more profound understanding of the intricate correlation between disease phenotype and the spatial arrangement of the tumor microenvironment.

\backmatter

\bmhead{Supplementary information}

This article includes supplementary material.

\clearpage
\newpage



\bibliography{sn-bibliography}

\clearpage
\newpage

\clearpage
\newpage

\begin{appendices}

\section{Appendix}\label{secA1}

\subsection{Datasets}
\label{sec:supp_dataset}

In this study, we used 6 datasets from 3 tissue types. The datasets included CoNSeP~\cite{graham2019hover}, NuCLS~\cite{amgad2021nucls}, PanNuke~\cite{gamper2020pannuke} (we refer to the breast and colon samples of this dataset as 2 different datasets), Lizard\cite{graham2021lizard}, and an internal dataset called \ovarianDatasetName dataset. The details of each dataset can be found in \Cref{dataset_tb}\par

\subsubsection{CoNSeP} consisted of 41 H\&E tiles from colorectal tissues extracted from 16 whole slide images of a single patient. All tiles were in a 40$\times$ magnification scale with the size of 1,000$\times$1,000 pixels. Cell types included 7 different categories including normal epithelial, malignant epithelial, inflammatory, endothelial, muscle, fibroblast, and miscellaneous. However, as suggested by the original paper, normal and malignant epithelial were grouped into the epithelial category, and the muscle, fibroblast, and endothelial cells were grouped into the spindle-shaped category. Therefore, the final 4 groups included epithelial, spindle-shaped, inflammatory, and miscellaneous. We followed the same train/test split of the original dataset. \par

\subsubsection{NuCLS} included 1744 tiles of breast cancer images from the TCGA dataset collected from 18 institutions. The tiles had different sizes, but they were roughly 300$\times$300 pixels. There were 12 different cell types available in this dataset: tumor, fibroblast, lymphocyte, plasma, macrophage, mitotic, vascular endothelium, myoepithelium, apoptotic body, neutrophil, ductal epithelium, and eosinophil. However, as suggested by the paper, these subtypes were grouped together into 5 superclasses including tumor (containing tumor and mitotic cells), stromal (containing fibroblast, vascular endothelium, and macrophage), sTILs (containing lymphocyte and plasma cells), apoptotic cells, and others. We used the first fold of this dataset for testing our model, while the rest of the dataset was used as the training data. \par

\subsubsection{PanNuke} included tiles from 19 different tissue types among which breast and colon were the most common ones. The slides were scanned at a 40$\times$ magnification scale, and cell types included neoplastic, epithelial, inflammatory, connective, and dead cells. In this study, we only focused on the breast and colon cells as they have a larger population compared to the rest of the tissues. Although the colon portion of the dataset included all 5 cell categories, the breast tissue does not contain any dead cells, leading to 4 cell types for this tissue. This dataset contained 3 folds in total; the first fold was used for training purposes, and the third one was adopted for testing. \par

\subsubsection{Lizard} included colon tissue tiles collected from 6 different sources across multiple sites in the world. The image tiles were scanned at the magnification scale of 20$\times$. Cells were divided into 6 categories including epithelial, connective tissue, lymphocyte, plasma, neutrophil, and eosinophil cells. This dataset contained 3 folds, of which the first and third were used for training and testing, respectively. \par

\subsubsection{\ovarianDatasetName} was an internal ovarian dataset collected in British Columbia, Canada. It included 192 TMA cores stained with H\&E alongside the multi-color IHC scans of an adjacent slice for each core. IHC images captured different biomarkers including CD3 (T-cell), CD94 (Natural Killer cell), FoxP3 (T-cell), CD79a (B-cell), CD8 (killer T-cell), CD68 (macrophages), CD16a (natural killer cell), and PanCK+ (epithelial cancer cells), each of which could be used as a map to find the type of cells in the corresponding TMA core. To this end, we registered each IHC image with its corresponding core. However, due to the circular shape of the images, only 11 of them were visually matched. The cells included in these images were used as the test set. \par

\subsubsection{SarcCell} was an internal soft tissue sarcoma dataset collected from epithelioid sarcoma cases. Whole tissue sections were obtained from formalin-fixed paraffin-embedded source blocks and stained with CD3 (T-cell, 1:100, Leica Biosystems), CD20 (B-cell, 1:200, Biocare), and CD208 (mature dendritic cells, 1:50,  Novus Biologicals) by multiplex. Adjacent sections were stained by H\&E. Stained sections were scanned at 40X and were co-registered together. IHC biomarkers were used as an indicator of the cell type for label generation. The distribution of extracted cells was extremely skewed towards T-cells (96\% of all the cells). Although all the extracted cells were used in the training set, we selected a balanced subset of the training set as the test set (including all the B-cells and dendritic cells, but a portion of the T-cells).

\begin{table}[h]
\begin{center}
    \caption{Datasets summary. The datasets are distributed across 4 tissue and 18 cell types to demonstrate the utility of the method. }
    \label{dataset_tb}
\adjustbox{max width=\textwidth}{
    \begin{tabular}{@{\extracolsep{\fill}}lccccp{4cm}c@{\extracolsep{\fill}}}
                                    Dataset           & Cell Type Count        & Tissue Type   & Magnification Scale   & Total Number of Cells & Cell Types & Number of Institutions\\
    \toprule
                                    CoNSeP            & 4                 & Colon         & $40\times$            & $24,319$              & Spindle, epithelial, inflammatory, miscellaneous & 1 \\ 
                                    \midrule
                                    NuCLS             & 5                 & Breast        & $20\times$            & $51,986$              & Tumor, stromal, sTIL, apoptotic, miscellaneous & 18 \\
                                    \midrule
                                    PanNuke Breast    & 4                 & Breast        & $40\times$            & $34,806$              & Neoplastic, epithelial, inflammatory, and connective & $>2$\\
                                    \midrule
                                    PanNuke Colon     & 5                 & Colon         & $40\times$            & $23,831$              & Neoplastic, epithelial, inflammatory, connective, and dead  & $>2$\\
                                    \midrule
                                    Lizard            & 6                 & Colon         & $20\times$            & $277,654$             & Epithelial, connective tissue, lymphocyte, plasma, neutrophil, and eosinophil  & $>7$\\
                                    \midrule
                                    \ovarianDatasetName            & 3                 & Ovarian       & $40\times$            & $265,580$             & Tumor, T-cell, B-cell, and Natural Killer & 1 \\
                                    \midrule
                                    SarcCell          & 3                 & Soft Tissue   & $40\times$            & $32,842$              & T-cell, B-cell, and dendritic & 1 \\
    \bottomrule
    \end{tabular}
    }    
\end{center}
\end{table}

\newpage
\subsection{Evaluation Metrics}
\label{sec:evaluation_metrics}

The clustering performance was measured based on multiple metrics, including Adjusted Mutual Information (AMI)~\cite{vinh2010information}, Adjusted Random Index (ARI)~\cite{hubert1985comparing}, and Purity of the identified cell clusters by the model and ground truth labels. In particular, AMI captures the agreement between two sets of assignments using mutual information while it is adjusted to mitigate the effect of chance on the score. On the other hand, ARI is the chance-adjusted form of the rand index~\cite{rand1971objective}, which calculates the quality of the clustering based on the number of matching instance pairs. Also, Purity measures how the samples within each cluster are similar to each other. In other words, it demonstrates whether each cluster is a mixture of different classes or not. \par

\newpage

\subsection{Results}\label{supp_results}

\subsubsection{Supervised Fine-tuning}

\begin{table}[h]
\begin{center}
\begin{minipage}{\textwidth}
    \caption{Fine-tuning accuracy for CoNSeP and NuCLS datasets. The supervised baselines demonstrate the performance of the HoVer-Net and NuCLS models on the CoNSeP and NuCLS datasets, respectively.}
    \label{finetuning_tb}
    \begin{tabular*}{\textwidth}{@{\extracolsep{\fill}}lcc@{\extracolsep{\fill}}}
    \toprule
                                            & CoNSeP            & NuCLS             \\
    \midrule
        Supervised Baseline                 & 79.6\%            & 77.5\%            \\
    \cmidrule{1-3}
        1-Layer MLP with frozen backbone    & 79.3\%            & 75.7\%            \\
        1-Layer MLP with unfrozen backbone  & 79.6\%            & 75.8\%            \\
        2-Layer MLP with frozen backbone    & 76.8\%            & 74.5\%            \\
        2-Layer MLP with unfrozen backbone  & \textbf{80.22\%}  & 76.3\%            \\
    \bottomrule
    \end{tabular*}
\end{minipage}    
\end{center}
\end{table}

\begin{table}[h]
\begin{center}
\begin{minipage}{\textwidth}
    \caption{Fine-tuning of NuCLS with color normalization}
    \label{nucls_finetuning_tb}
    \begin{tabular*}{\textwidth}{@{\extracolsep{\fill}}lc@{\extracolsep{\fill}}}
    \toprule
                                        & NuCLS             \\
    \midrule
    Fully-Supervised baseline           & 77.5\%            \\
    \midrule{1-2}
    1-Layer MLP with unfrozen backbone  & 78.1\%            \\
    2-Layer MLP with unfrozen backbone  & \textbf{78.2\%}   \\
    \bottomrule
    \end{tabular*}
\end{minipage}    
\end{center}
\end{table}

\subsubsection{Color Normalization}

Since the CoNSeP dataset contains the images from one patient and they were scanned in one institution, we excluded it from this experiment. 

\begin{table}[h]
\begin{center}
\begin{minipage}{\textwidth}
    \caption{Color normalization effect on unsupervised cell clustering}
    \label{normlization_tb}
    \begin{tabular*}{\textwidth}{@{\extracolsep{\fill}}lcccccc@{\extracolsep{\fill}}}
    \toprule
    &\multicolumn{3}{c}{w/o Color normalization} & \multicolumn{3}{c}{w/ Color normalization} \\
    \cmidrule{2-4}\cmidrule{5-7}%
         & AMI               & ARI               & Purity            & AMI                   & ARI               & Purity            \\
    \midrule
    \modelName & 26.1\%            & \textbf{25.6\%}   & 70.3\%            & \textbf{26.8}\%       & 22.9\%            & \textbf{70.8\%}   \\
    \bottomrule
    \end{tabular*}
\end{minipage}    
\end{center}
\end{table}

\subsubsection{Model Size Comparison}
Despite the fact that our model is capable of outperforming the supervised models, we noticed that it has 70\% fewer parameters compared to HoVer-Net. However, one might relate this matter to the fact that Hover-Net performs 2 tasks of cell detection and classification, simultaneously. To address this, we only included the parameters of the common encoder and the classification head of this model to have an impartial comparison (\Cref{model_size_tb}).

\begin{table}[h]
\begin{center}
\begin{minipage}{\textwidth}
    \caption{Model size comparison}
    \label{model_size_tb}
    \begin{tabular*}{\textwidth}{@{\extracolsep{\fill}}lc@{\extracolsep{\fill}}}
    \toprule
                                    & Inference Parameter Count \\
    \midrule
    HoVer-Net                       & 35.3M (common encoder: 25.6M, head: 9.7M) \\
    Mask-RCNN in NuCLS              & 130M \\
    Ours (1-Layer MLP)              & \textbf{11.17M} \\
    Ours (2-Layer MLP)              & \textbf{11.43M} \\
    \bottomrule
    \end{tabular*}
\end{minipage}    
\end{center}
\end{table}

\subsubsection{Cancer subtype classification}

To predict the subtype, the TMA core image was divided into $400\times400$ pixel patches, cells were divided into multiple clusters using the K-means algorithm, the number of cells within each predicted cluster was counted for each patch, and then patches were split into multiple groups based on the distributions of cell clusters. Afterward, we selected 100 patches from each group and predicted the clustering label of each original image based on the distribution of cell clusters in the selected patches. The final grouping process was conducted by using a hierarchical clustering algorithm with Ward's linkage method, preceded by dimensionality reduction using PCA.

\newpage
\subsection{Ablation Study}\label{sec:ablation_study}

To study the effects of each component of the framework, we also conducted multiple ablation studies. In this section, we limited the training duration to 200 epochs and reduced the batch size to 512, and the studies were performed only on the CoNSeP and NuCLS datasets to decrease the time and memory requirements. \par

\subsubsection{Memory Bank} \Cref{memeory_ablation_tb} shows the results of the unsupervised clustering performance in the presence and absence of the negative sample memory bank. These results confirmed that the presence of the memory bank is critical to the performance of the model. \par

Comparing the results of this table with that of \Cref{results_tb} (which was trained on 500 epochs with a batch size of 1024), we found that although the performance over the CoNSeP dataset dropped, the model produced almost similar results on the NuCLS dataset. We hypothesized that this must be related to the size of the dataset. Therefore, as the NuCLS dataset was 2$\times$ larger than the CoNSeP, the model could converge to a steady performance with even fewer training epochs. \par

\begin{table}[h]
\begin{center}
\begin{minipage}{\textwidth}
    \caption{Memory bank ablation study}
    \label{memeory_ablation_tb}
    \begin{tabular*}{\textwidth}{@{\extracolsep{\fill}}lcccccc@{\extracolsep{\fill}}}
    \toprule
    \multirow{2}{*}{}   & \multicolumn{3}{c}{CoNSeP}    & \multicolumn{3}{c}{NuCLS} \\
    \cmidrule{2-4}\cmidrule{5-7}%
                        & AMI               & ARI               & Purity            & AMI                   & ARI               & Purity            \\
    \midrule
    w/o Memory Bank     & 20.1\%            & \textbf {15.6\%}  & 57.8\%            & 22.5\%                & 24.4\%            & 66.6\%            \\
    w/ Memory Bank      & \textbf {21.3\%}  & 15.4\%            & \textbf {59.4\%}  & \textbf {26.82\%}     & \textbf {28.7\%}  & \textbf {69.35\%} \\
    \bottomrule
    \end{tabular*}
\end{minipage}    
\end{center}
\end{table}

\subsubsection{Masking cells in the environment patch} In this section, we studied the effects of the cell masking operation. In this regard, instead of masking all cells present in the environment patch, we only masked the target cell (the one that already exists in the associated single cell image fed into the Cell Block). The results (\Cref{masking_ablation_tb}) showed that the masking operation increased the general performance of the model, implying that reduced bias towards other cells’ types and more focus on the non-cellular environment such as tissue structure was important for environment integration. \par

\begin{table}[h]
\begin{center}
\begin{minipage}{\textwidth}
    \caption{Masking cells in the environment patch ablation study}
    \label{masking_ablation_tb}
    \begin{tabular*}{\textwidth}{@{\extracolsep{\fill}}lcccccc@{\extracolsep{\fill}}}
    \toprule
    \multirow{2}{*}{} & \multicolumn{3}{c}{CoNSeP} & \multicolumn{3}{c}{NuCLS} \\
    \cmidrule{2-4}\cmidrule{5-7}%
                        & AMI               & ARI               & Purity            & AMI                   & ARI               & Purity            \\
    \midrule
    w/o Masking     & 18.15\%           & 11\%              & 57\%              & 26.3\%                & 26.2\%            & \textbf{70.2\%}   \\
    w/ Masking      & \textbf{21.3\%}   & \textbf{15.4\%}   & \textbf{59.4\%}   & \textbf{26.82\%}      & \textbf{28.7\%}   & 69.35\%           \\
    \bottomrule
    \end{tabular*}
\end{minipage}    
\end{center}
\end{table}

\subsubsection{Multi-cropping} We compared the effect of multi-cropping (global-local cropping augmentations) with the conventional local-local relation where both of the augmentation pipelines in the Cell Block contained the cropping operation. Surprisingly, although multi-cropping showed a significant improvement on NuCLS dataset, it reduced the performance of the model on CoNSeP (\Cref{multicrop1_ablation_tb}). We hypothesized that this was due to the fact that the local augmentations obtained in this setting were not diverse enough for the model to be properly trained as the CoNSeP dataset had fewer data samples. Therefore, we increased the number of training epochs on the CoNSeP dataset and compared the effect of multi-cropping again. Our results (\Cref{multicrop2_ablation_tb}) demonstrated the positive effect of multi-cropping on the performance when the model was trained for a longer time. \par

\begin{table}[h]
\begin{center}
\begin{minipage}{\textwidth}
    \caption{Multi-cropping ablation study}
    \label{multicrop1_ablation_tb}
    \begin{tabular*}{\textwidth}{@{\extracolsep{\fill}}lcccccc@{\extracolsep{\fill}}}
    \toprule
    \multirow{2}{*}{} & \multicolumn{3}{c}{CoNSeP} & \multicolumn{3}{c}{NuCLS} \\
    \cmidrule{2-4}\cmidrule{5-7}%
                        & AMI               & ARI               & Purity            & AMI                   & ARI               & Purity            \\
    \midrule
    w/o Multi-crop     & \textbf{22.7\%}   & \textbf{21.4\%}   & \textbf{60.1\%}   & 23\%                  & 26.2\%            & 69\%              \\
    w/ Multi-crop      & 21.3\%            & 15.4\%            & 59.4\%            & \textbf{26.82\%}      & \textbf{28.7\%}   & \textbf{69.4\%}   \\
    \bottomrule
    \end{tabular*}
\end{minipage}    
\end{center}
\end{table}

\begin{table}[h]
\begin{center}
\begin{minipage}{\textwidth}
    \caption{Ablation study of multi-cropping on CoNSeP with longer epochs}
    \label{multicrop2_ablation_tb}
    \begin{tabular*}{\textwidth}{@{\extracolsep{\fill}}lccc@{\extracolsep{\fill}}}
    \toprule
    \multirow{2}{*}{} & \multicolumn{3}{c}{CoNSeP}  \\
    \cmidrule{2-4}
                        & AMI               & ARI               & Purity            \\
    \midrule
    w/o Multi-crop     & 22.7\%            & \textbf{20.24\%}  & 57.5\%            \\
    w/ Multi-crop      & \textbf{25.5\%}   & 19.3\%            & \textbf{63.2\%}   \\
    \bottomrule
    \end{tabular*}
\end{minipage}    
\end{center}
\end{table}

\subsubsection{Ensembling} We also compared the impact of using the trained momentum encoder instead of the backbone in the testing phase, and as \Cref{ensembling_ablation_tb} suggests, using the momentum encoder improved the performance of the model. This finding was expected since the momentum encoder equation forms a model similar to Polyak-Ruppert averaging~\cite{caron2021emerging}, aggregating the the encoder network weights across all training epochs. \par

\begin{table}[h]
\begin{center}
\begin{minipage}{\textwidth}
    \caption{Ablation study of ensembling}
    \label{ensembling_ablation_tb}
    \begin{tabular*}{\textwidth}{@{\extracolsep{\fill}}lcccccc@{\extracolsep{\fill}}}
    \toprule
    \multirow{2}{*}{} & \multicolumn{3}{c}{CoNSeP} & \multicolumn{3}{c}{NuCLS} \\
    \cmidrule{2-4}\cmidrule{5-7}%
                        & AMI               & ARI               & Purity            & AMI                   & ARI               & Purity            \\
    \midrule
    w/o Ensembling     & 20.7\%             & 13.5\%            & 58\%              & 24.4\%                & 25.2\%            & 69.3\%            \\
    w/ Ensembling      & \textbf{21.3\%}    & \textbf{15.4\%}   & \textbf{59.4\%}   & \textbf{26.82\%}      & \textbf{28.7\%}   & \textbf{69.4\%}   \\
    \bottomrule
    \end{tabular*}
\end{minipage}
\end{center}
\end{table}


\subsection{Cancer Subtype Clustering}\label{secA3}

\begin{figure}[p]
    \centering
    \includegraphics[width=0.8\textwidth]{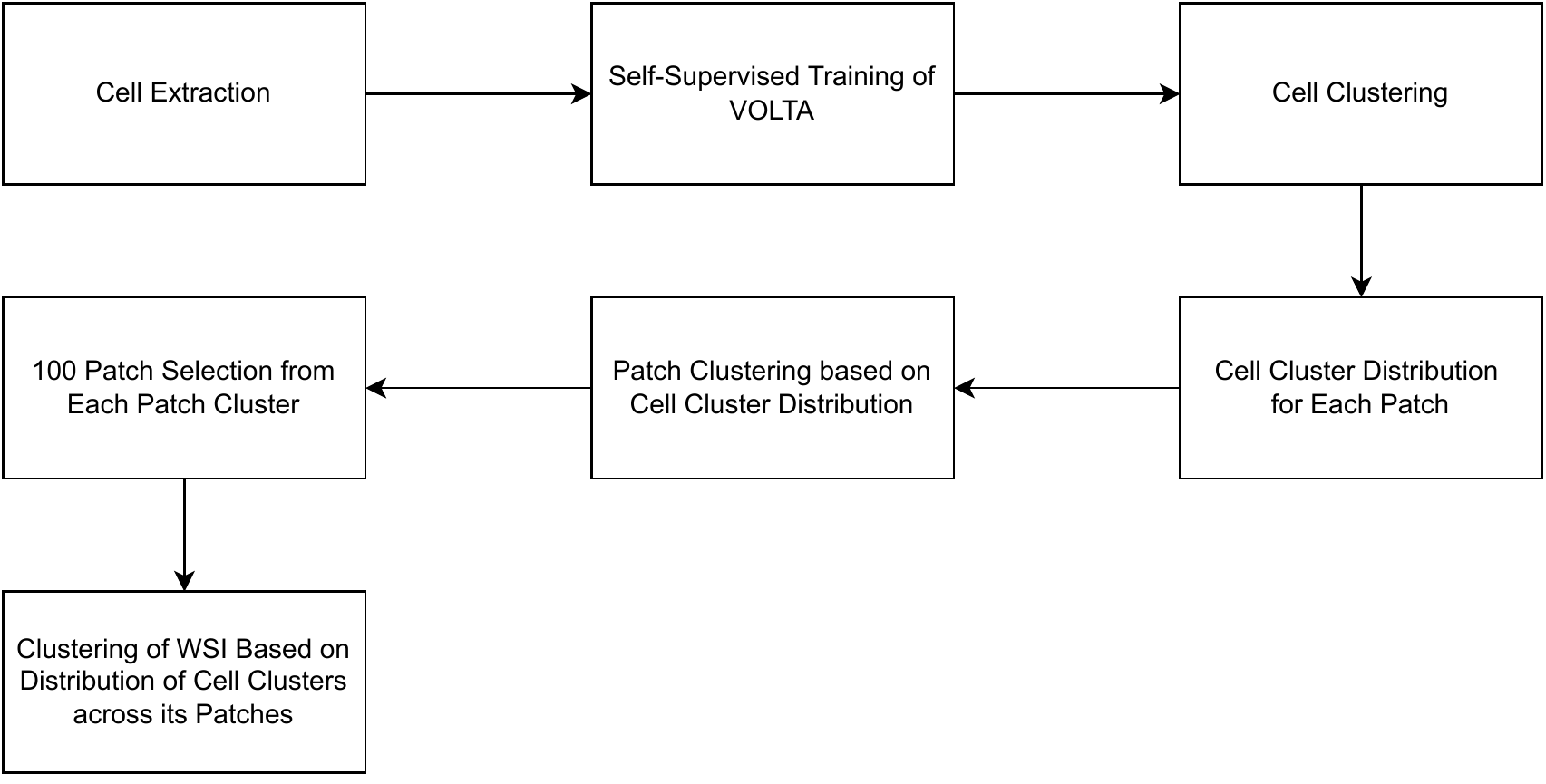}
    \caption{Workflow of unsupervised cancer subtype classification based on the unsupervised cell representation learning} 
    \label{cancer_subtyping_workflow_fg}
\end{figure}

\cref{cancer_subtyping_workflow_fg} demonstrates the flowchart of the process for cancer subtype classification on ovarian and endometrial datasets.

\begin{figure}[p]
    \centering
    \includegraphics[width=\textwidth]{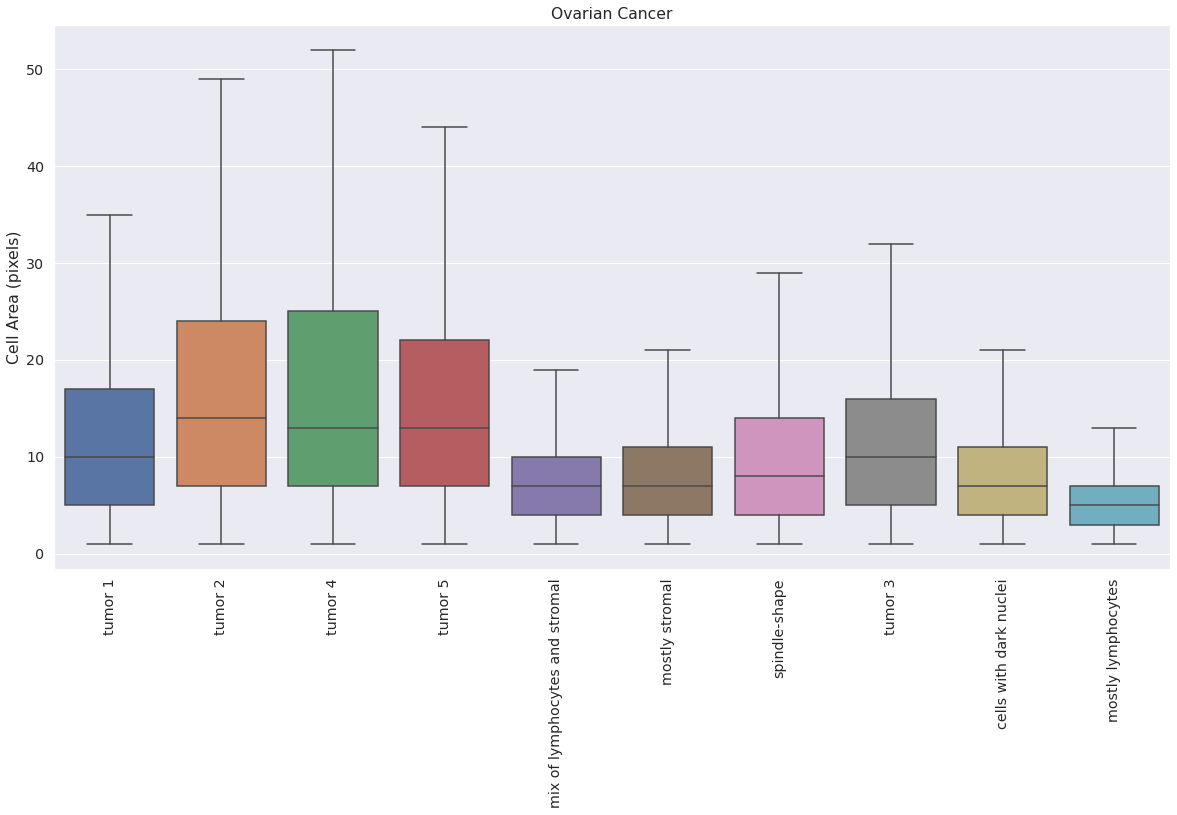}
    \caption{Cell area boxplot for cell clusters of the ovarian dataset} 
    \label{fig:ovarian_cell_area}
\end{figure}

\begin{figure}[p]
    \centering
    \includegraphics[width=\textwidth]{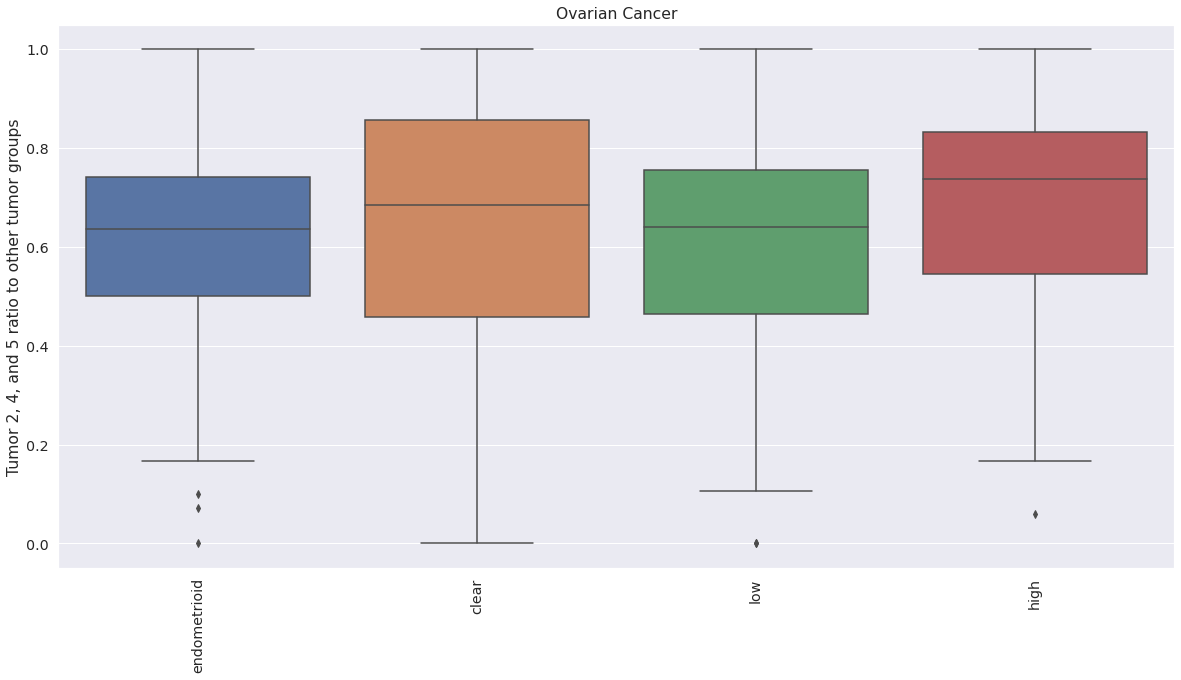}
    \caption{Tumor 2, 4, and 5 clusters proportion ratio with respect to all the tumor clusters for each histotype} 
    \label{fig:ovarian_tumor_245}
\end{figure}

\cref{fig:ovarian_cell_area} depicts the boxplot of the cell area for each cell clusters of the ovarian dataset, and \cref{fig:ovarian_tumor_245} shows the proportion of tumor 2, 4, and 5 clusters with respect to all the tumor clusters.

\begin{figure}[p]
    \centering
    \includegraphics[width=\textwidth]{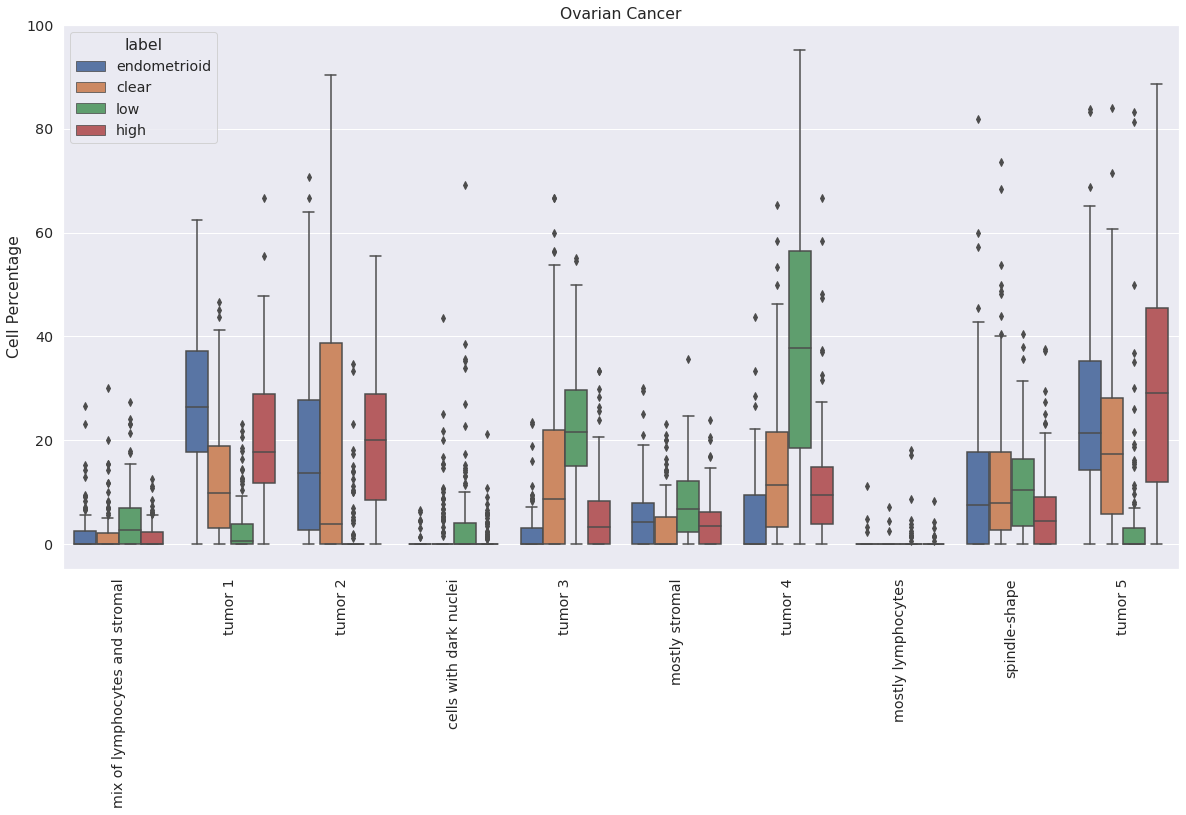}
    \caption{Boxplot of cell cluster distribution across patches of the ovarian cancer dataset for all of the patients. The blue, orange, green, and red boxes represent the percentage of each cell cluster for endometrioid, clear-cell, low-grade serous, and high-grade serous histotypes of ovarian cancer. } 
    \label{ovarian_boxplot_fg}
\end{figure}

\begin{table}[h]
\begin{center}
    \caption{Distribution of cell clusters across epithelial ovarian cancer histotypes (with standard deviation).}
    \label{ovarian_cluster_tb}
\adjustbox{max width=\textwidth}{
    \begin{tabular}{@{\extracolsep{\fill}}cccccccc@{\extracolsep{\fill}}}
    \toprule
           & Mixture of Lymphocytes and Small Stromal & Tumor  & Cells with Dark Nuclei & Stromal  & Lymphocytes  & Spindle-shape & Tumor and Stromal\\
    \midrule
    Clear Cell      & 3.0 $\pm 0.4\%$     & $59.5 \pm 1.0\%$     & $1.6 \pm 0.3\% $   & $8.8  \pm 0.6\%$     & $0.4 \pm 0.1\%$     & $12.5 \pm 0.8\%$    & $14   \pm 0.8 \%$ \\
    Endometrioid    & 3.3 $\pm 0.3\%$     & $57.2 \pm 0.9\%$     & $0.4 \pm 0.1\% $   & $10.6 \pm 0.6\%$     & $0.3 \pm 0.1\%$     & $0.3  \pm 0.6\%$    & $18.4 \pm 0.8 \%$ \\
    High grade      & 3.6 $\pm 0.3\%$     & $57.2 \pm 0.9\%$     & $1.9 \pm 0.2\% $   & $9.9  \pm 0.6\%$     & $0.5 \pm 0.1\%$     & $7.0  \pm 0.5\%$    & $19.6 \pm 0.8 \%$ \\
    Low grade       & 8.0 $\pm 0.4\%$     & $63.8 \pm 0.7\%$     & $3.5 \pm 0.3\% $   & $11.9 \pm 0.5\%$     & $1.6 \pm 0.1\%$     & $9.7  \pm 0.4\%$    & $1.0  \pm 0.2 \%$ \\
    Mucinous        & 1.3 $\pm 0.2\%$     & $54.7 \pm 0.9\%$     & $0.1 \pm 0.07\%$   & $6.9  \pm 0.5\%$     & $0.2 \pm 0.1\%$     & $6.5  \pm 0.4\%$    & $30   \pm 0.8 \%$ \\

    \bottomrule
    \end{tabular}
}
\end{center}
\end{table}

\begin{figure}[p]
    \centering
    \includegraphics[width=\textwidth]{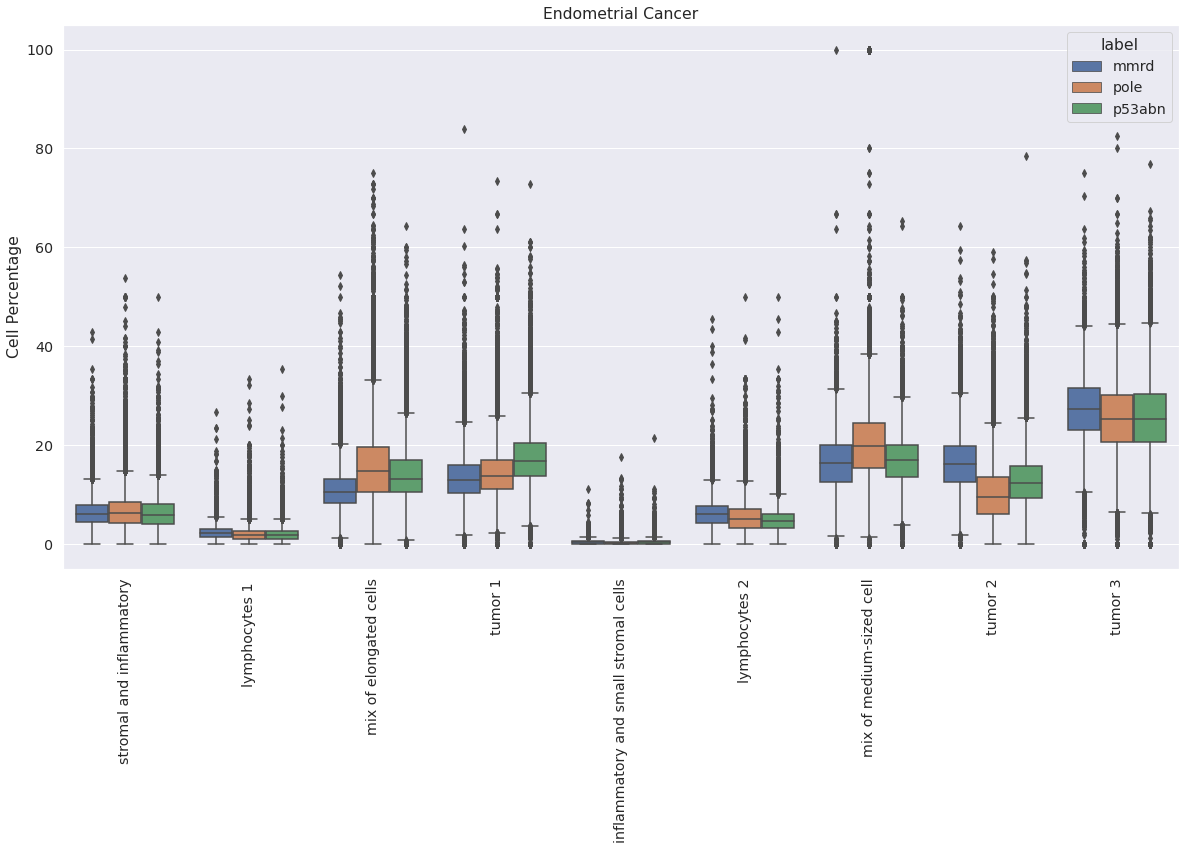}
    \caption{Boxplot of cell cluster distribution across patches of the Endometrial Cancer for all of the patients} 
    \label{endometrial_boxplot_fg}
\end{figure}

\begin{table}[h]
\begin{center}
    \caption{Distribution of cell clusters across endometrial cancer molecular subtypes (with standard deviation). The non-MMRd group encompasses p53abn and POLE tumors.}
    \label{endometrial_cluster_tb}
\adjustbox{max width=\textwidth}{
    \begin{tabular}{@{\extracolsep{\fill}}ccccccc@{\extracolsep{\fill}}}
    \toprule
           & Stromal and Inflammatory & Lymphocytes & Mixture of Elongated Cells & Tumor  & Inflammatory and Small Stromal & Mix of Medium-Sized Cells \\
    \midrule
    MMRd        & $6.8 \pm 0.03\%$     & $9.6 \pm 0.04\%$      & $10.3 \pm 0.04\%$      & $54.3 \pm 0.06\%$        & $0.4 \pm 0.008\%$     & $18.6 \pm 0.05\%$   \\
    non-MMRd    & $6.6 \pm 0.03\%$     & $7.9 \pm 0.03\%$      & $13.5 \pm 0.04\%$      & $51.0 \pm 0.06\%$        & $0.4 \pm 0.007\%$     & $20.6 \pm 0.04\%$  \\
    \midrule
    P53abn      & $6.3 \pm 0.2\% $     & $7.3 \pm 0.03\%$      & $12.9 \pm 0.04\%$      & $53.9 \pm 0.05\%$        & $0.4 \pm 0.007\%$     & $19 \pm 0.04\%$     \\
    POLE        & $6.8 \pm 0.03\%$     & $8.2 \pm 0.03\%$      & $13.8 \pm 0.04\%$      & $49.1 \pm 0.05\%$        & $0.3 \pm 0.007\%$     & $21.56 \pm 0.04\%$  \\
    \bottomrule
    \end{tabular}
}
\end{center}
\end{table}



\begin{figure}[p]
    \centering
    \includegraphics[width=\textwidth]{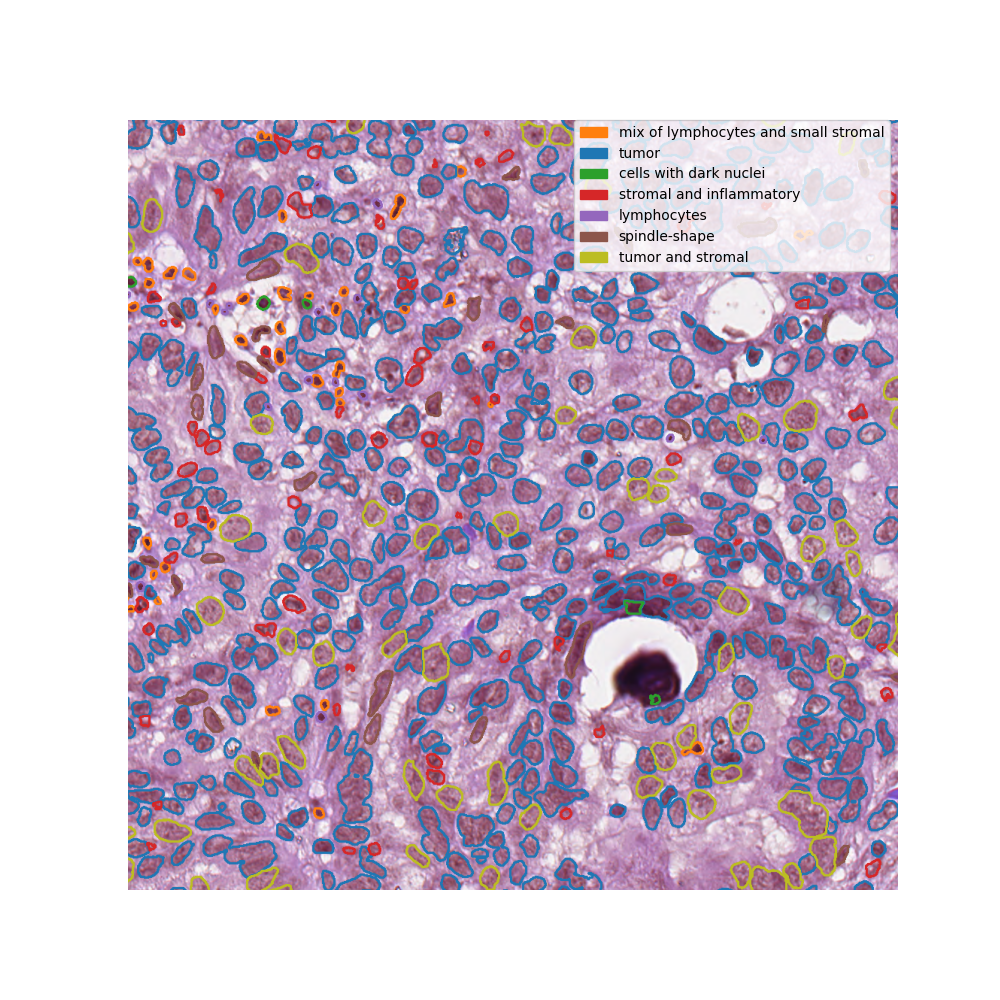}
    \caption{Inferred cell cluster labels in a representative example of a low-grade serous ovarian carcinoma with psammomatous calcification. Cell cluster labels are depicted as coloured nuclear outlines, and represent morphologically-distinct clusters that are associated with cell types. Green captures cells with darkly staining nuclei, and brown labels spindled cells that are mostly stromal.} 
    \label{ovarian_low_grade_overlay_fg}
\end{figure}

\begin{figure}[p]
    \centering
    \includegraphics[width=\textwidth]{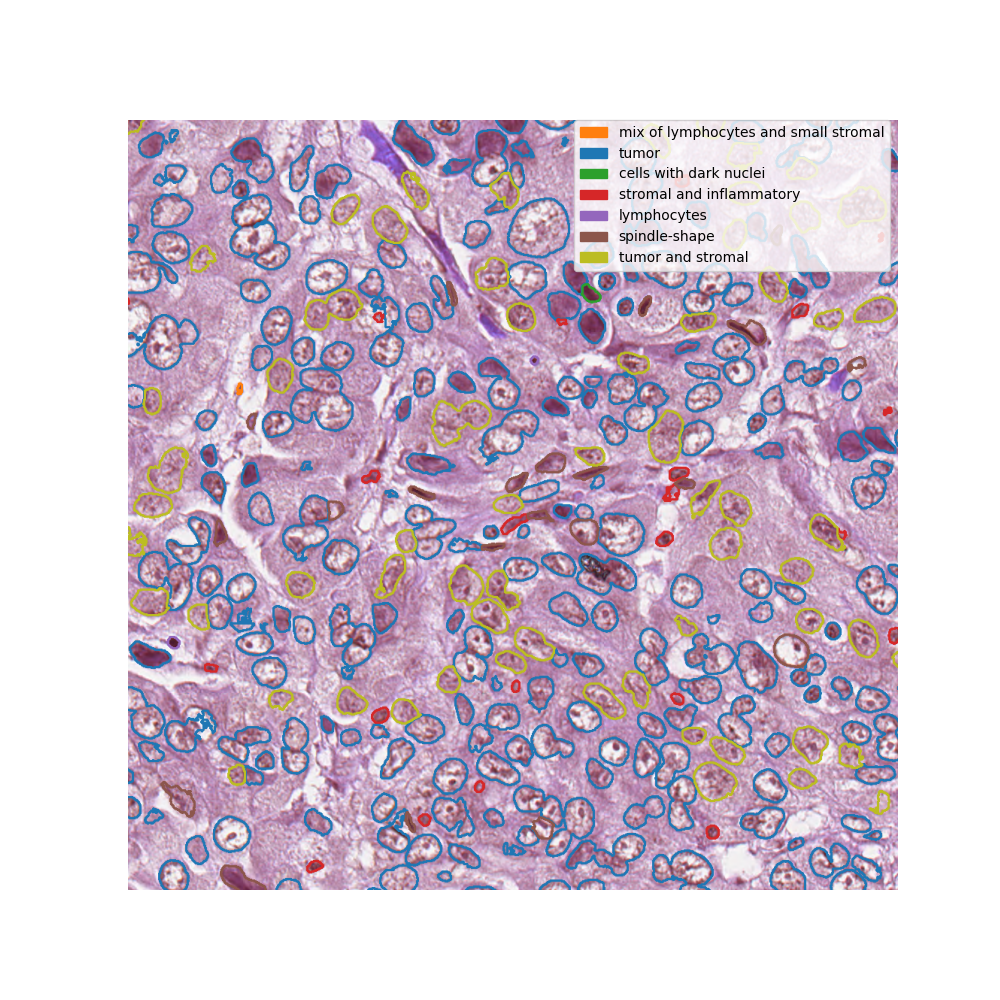}
    \caption{Inferred cell cluster labels in a representative example of a high-grade serous ovarian carcinoma. Cell cluster labels are depicted as coloured nuclear outlines, and represent morphologically-distinct clusters that are associated with cell types. Green captures cells with darkly staining nuclei, and brown labels spindled cells that are mostly stromal.} 
    \label{ovarian_high_grade_overlay_fg}
\end{figure}

\begin{figure}[p]
    \centering
    \includegraphics[width=\textwidth]{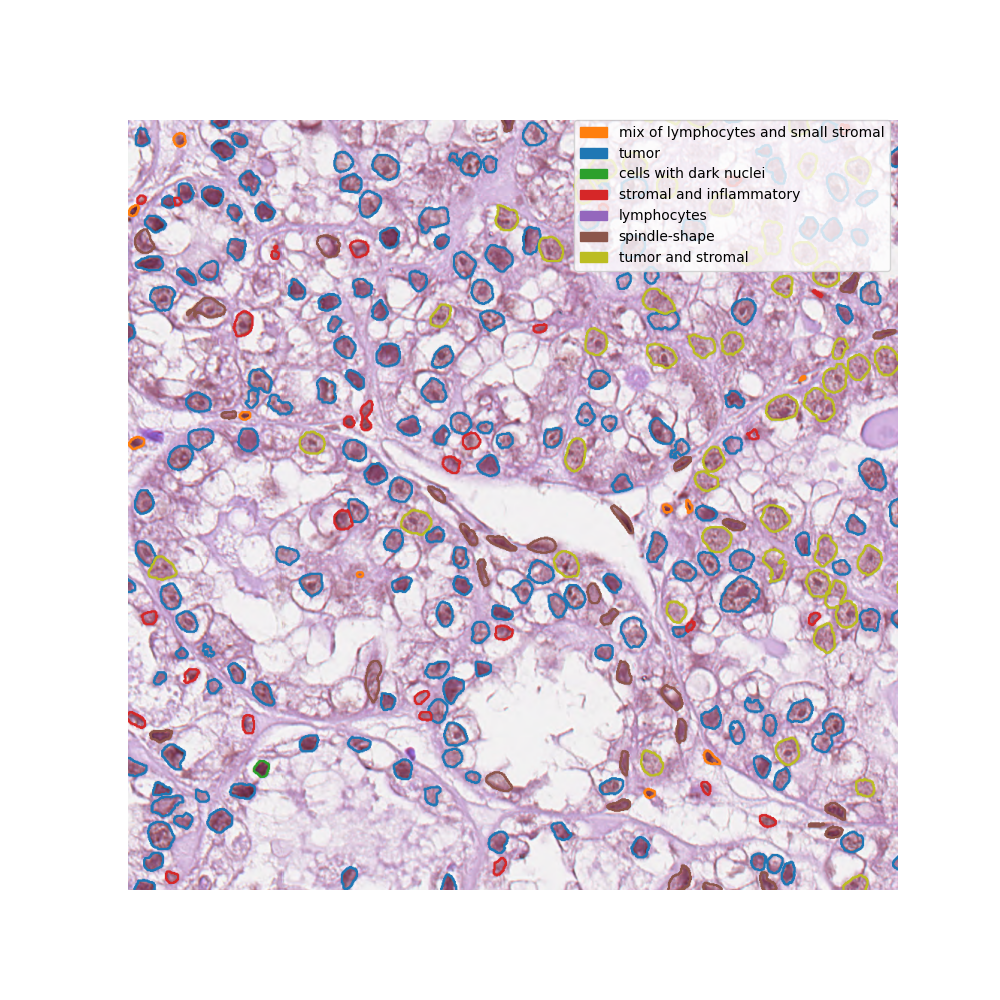}
    \caption{Ovarian - clear cell overlay visualization} 
    \label{ovarian_clear_overlay_fg}
\end{figure}

\begin{figure}[p]
    \centering
    \includegraphics[width=\textwidth]{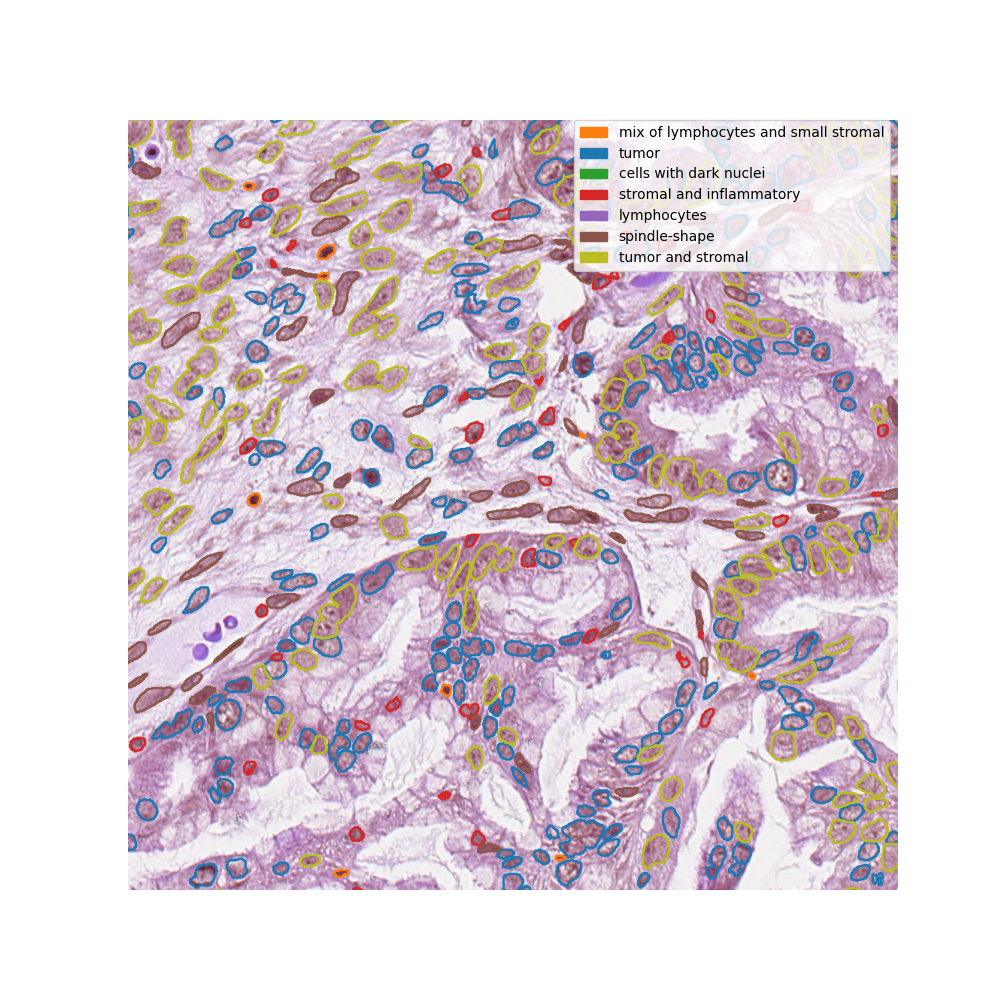}
    \caption{Inferred cell cluster labels in a representative example of a mucinous ovarian carcinoma. Cell cluster labels are depicted as coloured nuclear outlines, and represent morphologically-distinct clusters that are associated with cell types. Green captures cells with darkly staining nuclei, and brown labels spindled cells that are mostly stromal. In this example, blue also captures reactive stromal cells with plump nuclei.} 
    \label{ovarian_mucinous_overlay_fg}
\end{figure}

\begin{figure}[p]
    \centering
    \includegraphics[width=\textwidth]{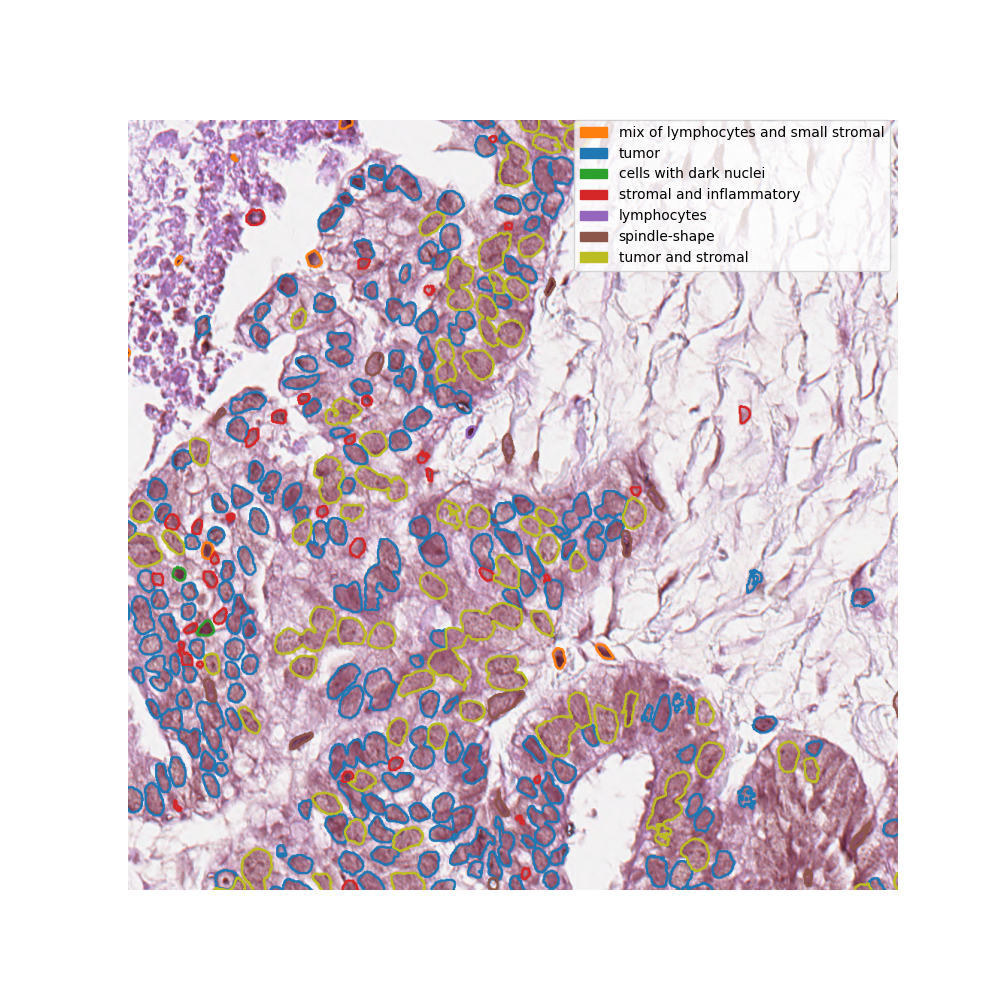}
    \caption{Inferred cell cluster labels in a representative example of an endometrioid ovarian carcinoma. Cell cluster labels are depicted as coloured nuclear outlines, and represent morphologically-distinct clusters that are associated with cell types. Green captures cells with darkly staining nuclei, and brown labels spindled cells that are mostly stromal.} 
    \label{ovarian_endometriod_overlay_fg}
\end{figure}

\begin{figure}[p]
    \centering
    \includegraphics[width=\textwidth]{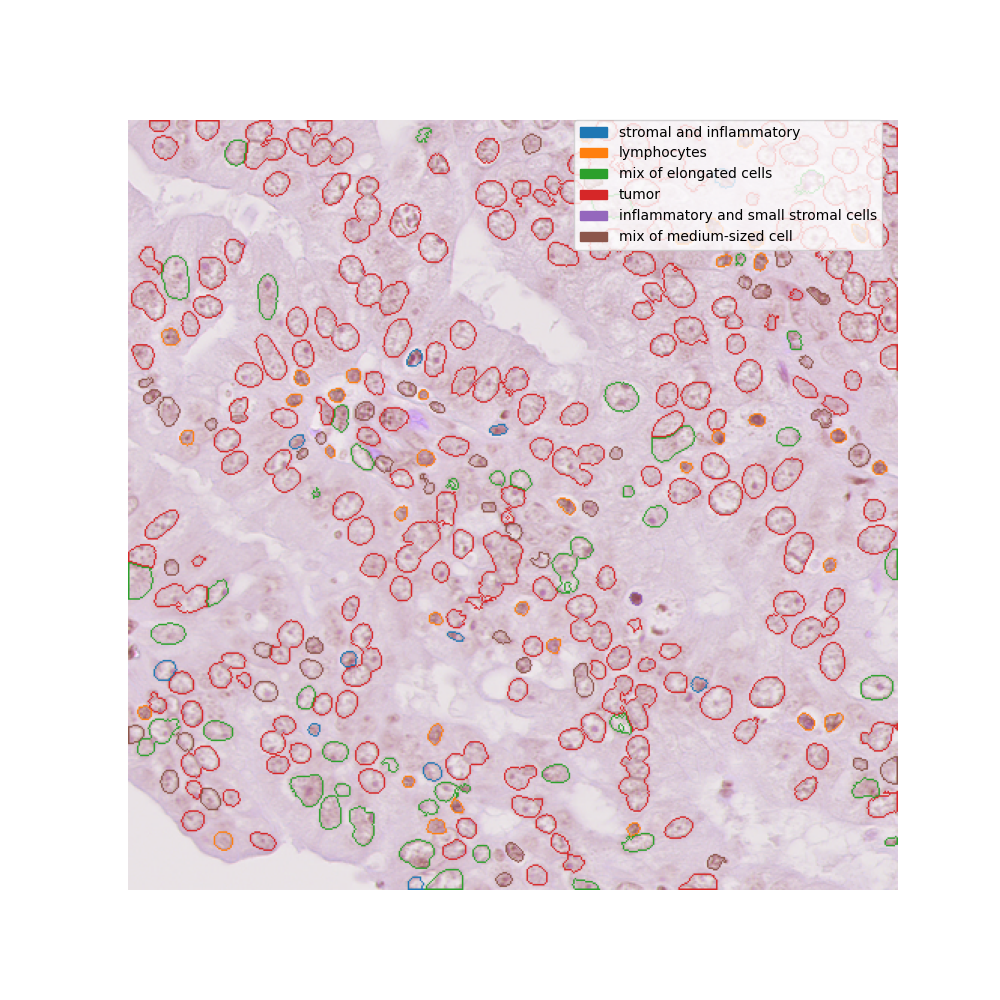}
    \caption{Inferred cell cluster labels in a representative example of a POLE-mutated endometrial carcinoma. Cell cluster labels are depicted as colored nuclear outlines, and represent morphologically-distinct clusters that are associated with cell types. Green cells comprise a mixture of stromal and cancer cells with elongated nuclei. Brown denotes cells with intermediate-sized nuclei, comprising a mixture of tumor, inflammatory, and stromal cells. In this example, some cancer cells are labeled green, highlighting the uncertainty of cell type classification in morphologically pleomorphic POLE tumors.} 
    \label{endometrial_pole_overlay_fg}
\end{figure}

\begin{figure}[p]
    \centering
    \includegraphics[width=\textwidth]{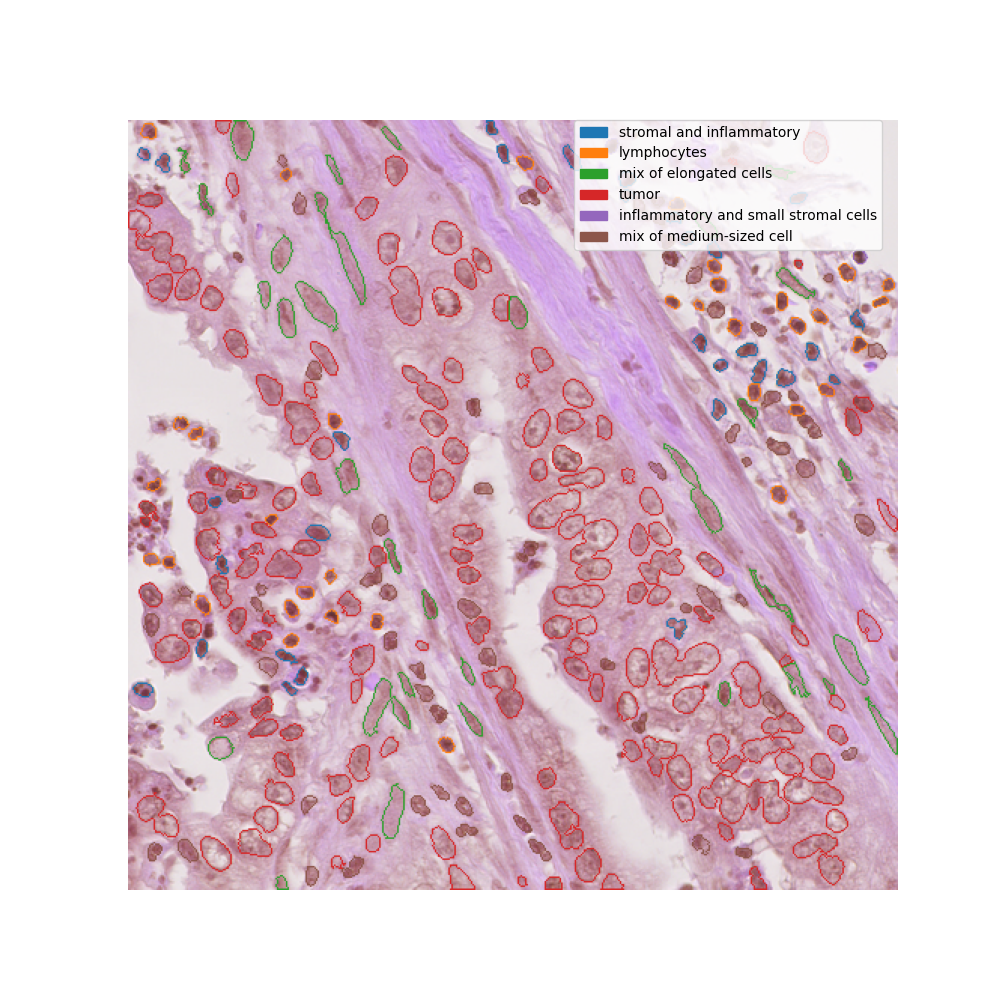}
    \caption{Endometrial - MMRd overlay visualization} 
    \label{endometrial_mmrd_overlay_fg}
\end{figure}

\begin{figure}[p]
    \centering
    \includegraphics[width=\textwidth]{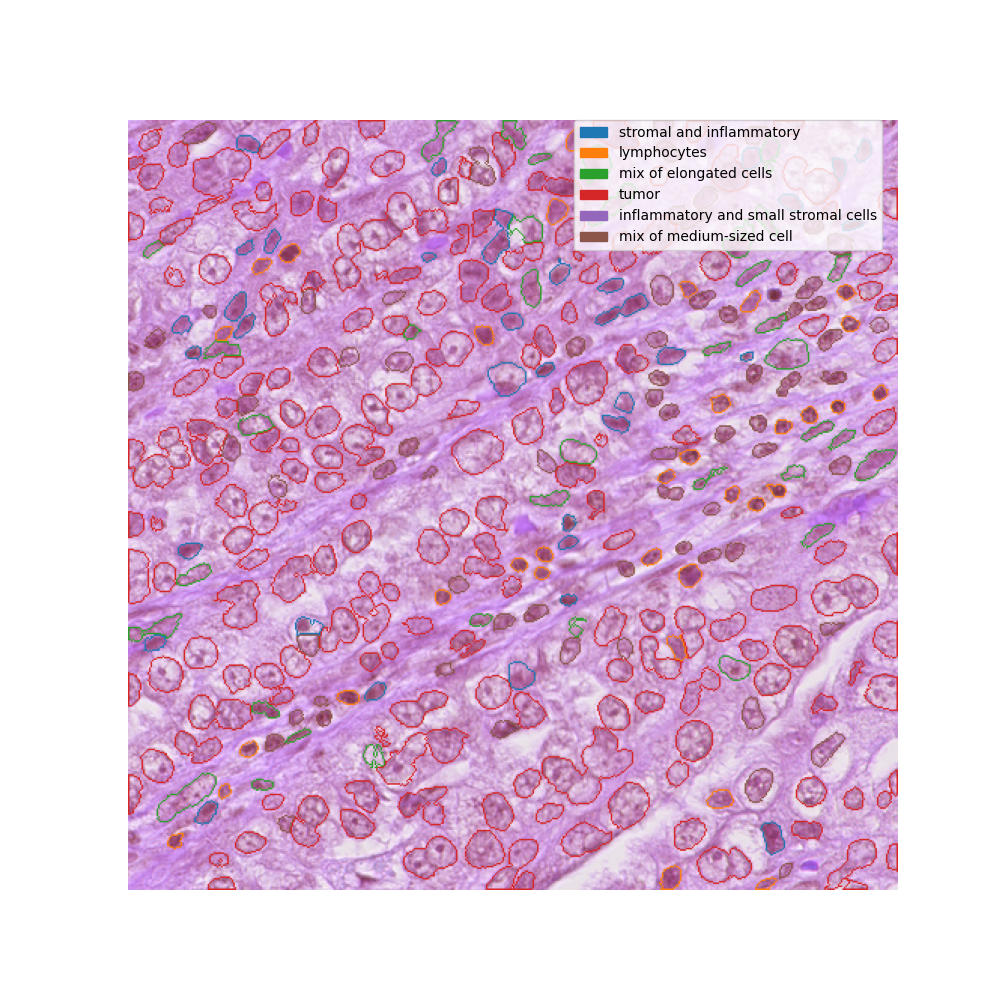}
    \caption{Inferred cell cluster labels in a representative example of a p53-mutated endometrial carcinoma. Cell cluster labels are depicted as coloured nuclear outlines, and represent morphologically-distinct clusters that are associated with cell types. Green cells comprise mixture of stromal and tumour cells with elongated nuclei. Brown denotes cells with intermediate sized nuclei, comprising a mixture of tumour, inflammatory, and stromal cells.} 
    \label{endometrial_p53abn_overlay_fg}
\end{figure}




\end{appendices}


\end{document}